\documentclass{ws-ijam}
\usepackage{graphics}
\usepackage{graphicx}
\usepackage{float}
\usepackage{subfigure}

\begin{document}

\markboth{Omer San and Anne E. Staples}
{Dynamics of pulsatile flows through elastic microtubes}

\catchline{}{}{}{}{}

\title{DYNAMICS OF PULSATILE FLOWS THROUGH ELASTIC MICROTUBES}

\author{OMER SAN\footnote{Corresponding author address: Department of Engineering Science and Mechanics, Virginia Tech, Blacksburg, VA 24061, USA. E-mail: omersan@vt.edu, Phone: +1 (540) 231 7482.}}

\address{
Department of Engineering Science and Mechanics\\
Virginia Tech,
Blacksburg, VA 24061, USA}

\author{ANNE E. STAPLES}  

\address{Department of Engineering Science and Mechanics\\
Virginia Tech,
Blacksburg, VA 24061, USA}

\maketitle


\begin{abstract}
We investigate the dynamics of pressure driven transient flows of incompressible Newtonian fluids through circular microtubes having thin elastic walls under the long-wavelength and small deformation assumptions, which are valid for many industrial and biological processes. An analytical solution of the coupled fluid and solid equations is found using Navier slip boundary conditions and is shown to include some existing Womersley solutions as limiting cases. The effect of the slip length at the fluid-solid interface is analyzed for oscillatory pressure gradients using a range of slip ratio and frequency parameters. The solutions for elastic and rigid walls are compared for the cases with and without slip boundary conditions for a broad range of the relevant parameters. It is shown that the elastic behavior of the microtube couples nonlinearly with the slip velocity, which greatly enhances the achievable flow rate and pumping efficacy compared to the inelastic case. In addition, it is observed that increasing the slip length produces less shear stress, which is consistent with the nearly frictionless interfaces observed in many microscale experiments.

\end{abstract}

\section{Introduction}
\label{sec:intro}
Understanding internal fluid flows at small scales is currently a subject of great interest for the development of biological and engineering devices and systems at the micro- and nanoscales [Beebe {\it et al.}, 2002;  Squires and Quake, 2005]. Since the flow dynamics determine the characteristics of the systems, the study of micro and nanoflows is of vital importance to developing better technologies. The behavior of pulsatile flows at the micro- and nanoscales, in particular, is a largely unexplored area [Hansen and Ottesen, 2006; Vedel {\it et al.}, 2010], and a better understanding of such flows is needed for systems with these characteristics.

Studies of fluid behavior at the micro- and nanoscales has resulted in the growing understanding that fluid mechanics at such small scales differs in fundamental ways from fluid mechanics in the macroscopic world [Gad-el Hak, 1999]. One of the fundamental assumptions in fluid mechanics is the no-slip boundary condition: that the tangential component of the fluid velocity equals that of the solid at the wall. Recent advances in micro- and nanoscale experiments and molecular simulations have shown that slip of the fluid on the solid surface occurs at small scales, and thus the traditional no-slip boundary condition cannot be applied at these scales [Barrat and Bocquet, 1999; Baudry {\it et al.}, 2001; Choi {\it et al.}, 2003; Joseph and Tabeling, 2005; Thompson and Robbins, 1990; Thompson and Troian, 1997]. Many other experimental and theoretical studies demonstrating the slip phenomenon have been recently reviewed by Whitby and Quirke [2007] and Rothstein [2010], along with other small scale fluid dynamics phenomena such as surface roughness, non-wettability, dissolved molecules, corrugation and altered shear rate. Neto {\it et al.} [2005] explained the most evident mechanisms by which both true slip (where fluid molecules are effectively sliding on the solid surface) and apparent slip (where the slip occurs not at the solid/fluid interface but in the thin depleted fluid/fluid interface layer near the solid boundary). For water over hydrophobic surfaces, or for any liquid for which the surface is less attractive than the bulk of the fluid (i.e., a non-wetting fluid-solid interface), there is an especially large slip at the fluid-solid interface with corresponding drag reduction properties, that is promising for a wide range of fluid dynamics and heat transfer applications such as liquid cooling for microelectronics [Ou {\it et al.}, 2004; Rothstein 2010, Watanabe {\it et al.} 1999].

Whether a fluid is a liquid or gas the magnitude of the slip at the wall can be quantified by the slip length, $\ell_s$, at solid surfaces [Thompson and Troian, 1997]. Nearly two centuries ago Navier proposed a generalized boundary condition that incorporates the possibility of fluid slip at the solid surface. It assumes that the slip velocity is proportional to the shear stress at the wall via the slip length. Its equivalent form for gases agrees with the kinetic theory of gases, which equates the slip length with the mean free path of the gas. One of the most important dimensionless physical parameters for fluid flows through microtubes is defined as $\lambda=\ell_s/R$ where $R$ is the radius of the tube (it is called the Knudsen number for gases, and in that case $\ell_s$ is the mean free path). Both gas and liquid flow measurements show dramatic enhancements in flux rates over those predicted with classical continuum no-slip flow models [Neto {\it et al.} 2005; Whitby and Quirke, 2007]. Majumder {\it et al.} [2005] observed that the measured flow rates through carbon nanotubes were four to five orders of magnitude greater than what was predicted by hydrodynamics models. The authors also showed that the slip lengths, which are much greater than the tube diameter, are consistent with a nearly frictionless interface due to the lower molecular corrugation. Similar observations were reported by Cheng and Gordanio [2002].

Experiments overwhelmingly show that the presence of a slip velocity on the boundary can play an important role in small scale fluid devices. Theoretical solutions to various flow problems for Newtonian fluids with slip boundary conditions have been obtained and are available in the literature [Karniadakis {\it et al.}, 2005]. Fluid flows through rigid microtubes have been studied in the slip flow regime [Sbragaglia and Prosperetti, 2007], and this study determined that an effective Navier slip boundary condition is valid. Other studies have investigated steady-state [Matthews and Hill, 2007] and transient [Wu {\it et al.}, 2008] flows through rigid microtubes with slip boundary conditions. Wu {\it et al.} [2008] showed that the influence of boundary-slip on the flow behavior is different for different types of pressure gradients driving the flow. Since there is no wave motion in the rigid tubes, the fluids oscillate in the bulk, and hence the effect of wave propagation was not modeled in those studies.

Unsteady flow characteristics and wave propagation through elastic tubes has been studied on the macroscale [Wormersly, 1955; Zamir, 2000]. There is no corresponding study in the literature, however, for the micro- or nanoscale. Pulsatile microfluidics is a widely unexplored area and the difference in the dynamics at the macro- and microscales remains to be quantified. Pulsatile microfluidics through the elastic conduits is relevant to many microsystems. For example, the boundary tuning mechanism in surface acoustic wave driven flows (for precise control of fluids through a miniaturized network) [Chu, 2004]. Another relevant application is valveless pumping mechanisms. Thus, a study of fluid flows in deformable micro- or nanotubes should be performed to better understand some natural process such as respiratory mechanism in insects [Westneat {\it et al.}, 2003], flow characteristics in micro- and nanofluidic systems, and the efficient design of these systems. We undertake such a study here. In particular, we derive an analytical solution for transient flows through elastic microtubes with Navier slip boundary conditions. The nonlinear influence of the slip length on the material deformation of the fluid significantly affects the velocity profiles, flow rates, and shear stress distributions in the system. Our results may be useful to researchers in the microfluidic device and biology communities for understanding the effects of pulsatile flow dynamics, elastic conduits, and slip velocities in internal flows at small scales.

\section{Mathematical formulation}
\label{sec:Math}
\begin{figure}[H]
\centering
\includegraphics[width=0.5\textwidth]{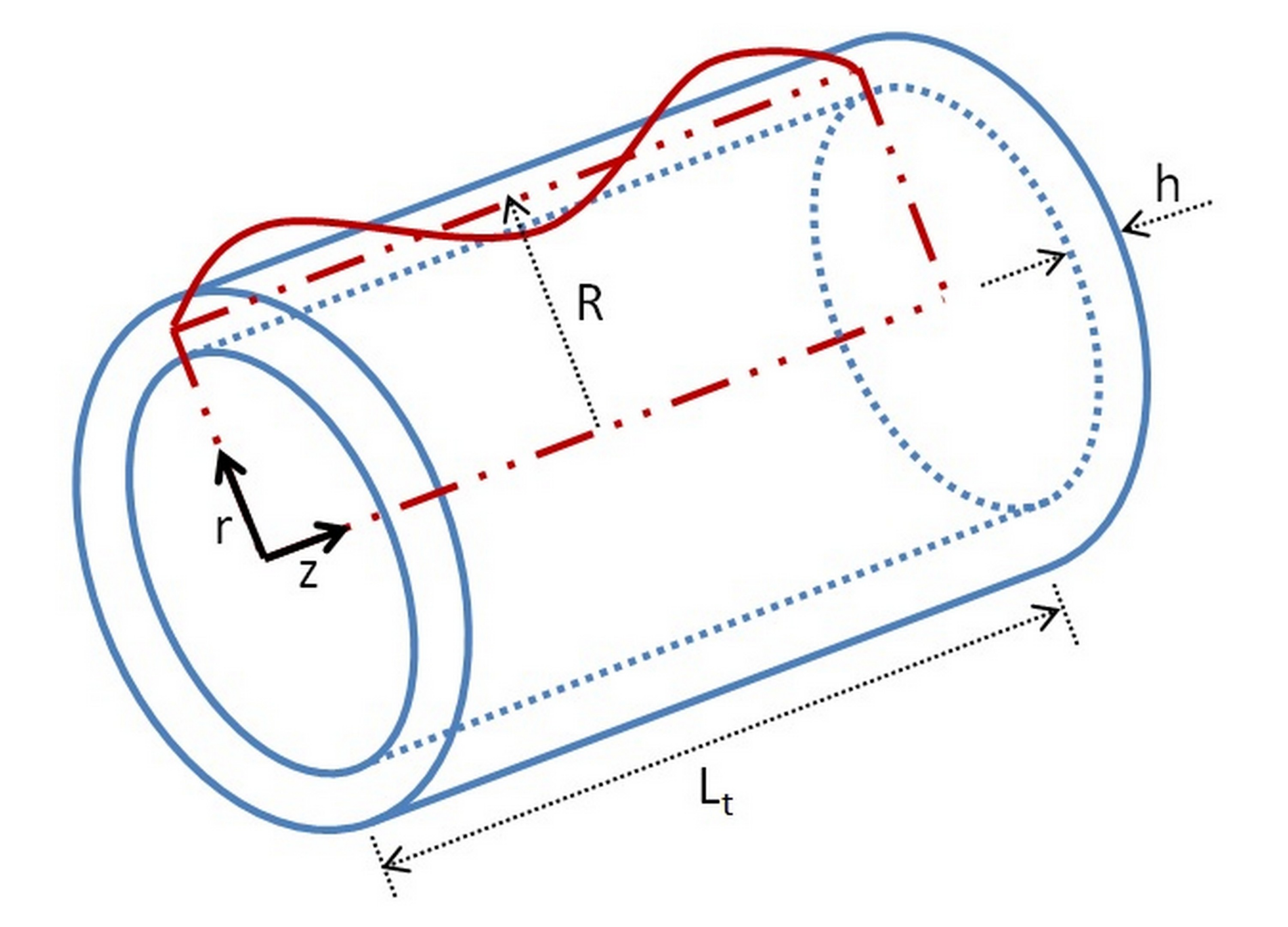}
\caption{Schematic of elastic microtube with cylindrical coordinates}
\label{fig:tube}       
\end{figure}

We consider a circular microtube with a thin elastic wall filled with a incompressible Newtonian fluid as illustrated in Fig.~\ref{fig:tube}. The governing equations for the conservation of momentum of incompressible flows can be written in vector form as
\begin{equation}
\frac{\partial \textbf{u}}{\partial t} +  \textbf{u} \cdot \nabla \textbf{u}  =-\frac{1}{\rho}\nabla p + \frac{\mu}{\rho} \nabla^2 \textbf{u} + f
\end{equation}
where $\mu$ and $\rho$ are the viscosity and density of the fluid, assumed constant for incompressible Newtonian fluids, $\textbf{u}$ and $p$ are the velocity and pressure fields of the fluid, and $f$ represents external body forces.
These equations can be simplified according to the conditions for which we would like to solve them.  The assumptions we make in this study are: (i) the wave speed of propagation is assumed to be much greater than the average flow velocity (which simplifies the nonlinear terms in the Navier-Stokes equation), (ii) the length of tube is much greater than the tube radius (which simplifies some viscous terms having an axial derivative), (iii) there is no external force such as gravity, (iv) the flow is axisymmetric and there is no swirling flow (no dependence on angular direction), (v) the thickness of the tube wall is much smaller than the tube radius (membrane equations for thin-walled elastic structures may be used), (vi) linearized coupling boundary conditions at the fluid-structure interface (which eliminates the necessity of a domain transformation), and (vii) the axial fluid velocity relative to the solid surface is proportional to the shear stress on the interface (Navier slip boundary conditions). These assumptions are valid in many biological and industrial processes.
\subsection{Governing equations for the fluid}
\label{sec:fluid}
Under assumptions (i-iv) the axial and radial momentum equations, and the continuity equation for an incompressible Newtonian fluid in cylindrical coordinates become [Womersley, 1955; Zamir, 2000]
\begin{equation}
\frac{\partial u}{\partial t}=-\frac{1}{\rho}\frac{\partial p}{\partial z} + \frac{\mu}{\rho}(\frac{\partial^2 u}{\partial r^2} + \frac{1}{r}\frac{\partial u}{\partial r})
\end{equation}
\begin{equation}
\frac{\partial v}{\partial t}=-\frac{1}{\rho}\frac{\partial p}{\partial r} + \frac{\mu}{\rho}(\frac{\partial^2 v}{\partial r^2} + \frac{1}{r}\frac{\partial v}{\partial r} - \frac{v}{r^2})
\end{equation}
\begin{equation}\label{con}
\frac{\partial u}{\partial z}+ \frac{\partial v}{\partial r} + \frac{v}{r} = 0
\end{equation}
where $p$ is the fluid pressure (since only the gradient of the pressure occurs in the equation, it is usually called gage or transmural pressure) and $u$, $v$ are the axial and radial velocity components. The stress field in the incompressible Newtonian fluid is related to the velocity field by the following constitutive relation
\begin{equation}\label{con}
\sigma = -p {\bf I} + 2\mu {\bf d}
\end{equation}
where the shear stress is defined as $\tau = 2 \mu {\bf d}$ while the rate of deformation tensor {\bf d} is related to the velocity vector by
\begin{equation}\label{con}
{\bf d} = \frac{1}{2}(\nabla \vec{u} + (\nabla \vec{u})^T)
\end{equation}
where $\vec{u} = v \hat{e}_r + u \hat{e}_z$ and therefore the velocity gradient tensor becomes:
\begin{equation}
\nabla \vec{u}=
\left[\begin{array}{ccc}
\frac{\partial v}{\partial r} & 0 & \frac{\partial u}{\partial r} \\
0 & \frac{v}{r} & 0 \\
\frac{\partial v}{\partial z} & 0 & \frac{\partial u}{\partial z}
\end{array} \right]
\label{eq:sys}
\end{equation}
The shear stress can then be written as
\begin{equation}
\tau_{rz}=\mu (\frac{\partial u}{\partial r} + \frac{\partial v}{\partial z})\approx \mu \frac{\partial u}{\partial r}
\end{equation}
\subsection{Governing equations for the tube wall}
\label{sec:wall}
Using assumption (v), for thin-walled elastic circular shells the elasticity equations for radial and axial displacements of the tube wall are given by [Womersley, 1955; Zamir, 2000]
\begin{equation}\label{con}
\frac{\partial^2 \xi}{\partial t^2} =\frac{E}{(1-\nu^2) \rho_{w}}(\frac{\partial^2 \xi}{\partial z^2} + \frac{\nu}{R}\frac{\partial \eta}{\partial z} ) + \frac{\tau_{w}}{h \rho_w}
\end{equation}
\begin{equation}\label{con}
\frac{\partial^2 \eta}{\partial t^2} =\frac{p_w-p_0}{h \rho_w} - \frac{E}{(1-\nu^2) R \rho_{w}}(\frac{\eta}{R} + \nu\frac{\partial \xi}{\partial z} )
\end{equation}
where $p_0$ and $p_w$  are the values of the pressure on the outer surface and inner surface of the tube, respectively. Here, $R$ and $h$ are the reference radius and the thickness of the tube as shown in Fig.~\ref{fig:tube}, and $\rho_{w}$, $E$, and $\nu$ are the density, Young's modulus, and Poisson ratio of the tube material.
\subsection{Fluid-structure coupling at the interface}
\label{sec:fsi}
The dynamic forces from the fluid acting on the structure are modeled in terms of the shear stress in the axial direction and the pressure in the radial direction and give rise to the dynamic coupling conditions
\begin{equation}
p_w-p_0 = p.
\end{equation}
\begin{equation}
\tau_w = -\tau_{rz}
\end{equation}
where $p$ is the transmural pressure. The kinematic boundary conditions are
\begin{equation}
\frac{\partial \xi}{\partial t}(z,t) = u_w(z,t)
\end{equation}
\begin{equation}
\frac{\partial \eta}{\partial t}(z,t) = v_w(z,t),
\end{equation}
where $u_w$ and $v_w$ are the axial and radial velocities of tube. Here, we linearize the actual boundary conditions and apply them only approximately at the neutral position of the wall, $r=R$, using assumption (vi).
\subsection{Navier slip boundary conditions}
\label{sec:slip}
Recent experiments and molecular simulations in microfluidics confirm that slip of fluid on solid surfaces occurs at small scales [Whitby and Quirke, 2007]. In this study, the generalized nonlinear Navier boundary condition advocated by Thompson and Troian [1997] is applied to the conventional continuum mechanical description of fluid flow considered here. The standard no-slip boundary condition is replaced by the nonlinear Navier slip boundary condition, wherein the slip velocity is assumed to be proportional to the tangential viscous stress and the degree of slip is measured by a non-constant slip length, $\ell_s$. I.e., the axial fluid velocity relative to the tube wall is linearly proportional (by assumption (iv)) to the shear stress at the wall. The kinematic matching condition for the axial velocity can then be written using the Navier slip boundary condition as
\begin{equation}
u(R,z,t) - u_w(z,t) =-\ell_s \frac{\sigma_{rz}(R,z,t)}{\mu}
\end{equation}
giving
\begin{equation}
u(R,z,t) - u_w(z,t) = -\ell_s \frac{\partial u}{\partial r}(R,z,t)
\end{equation}
and the kinematic matching conditions for the radial velocity component are
\begin{equation}
v(R,z,t) - v_w(z,t) =0.
\end{equation}

\section{Exact solutions}
\label{sec:Sol}
\begin{figure}
\centering
\includegraphics[width=0.75\textwidth]{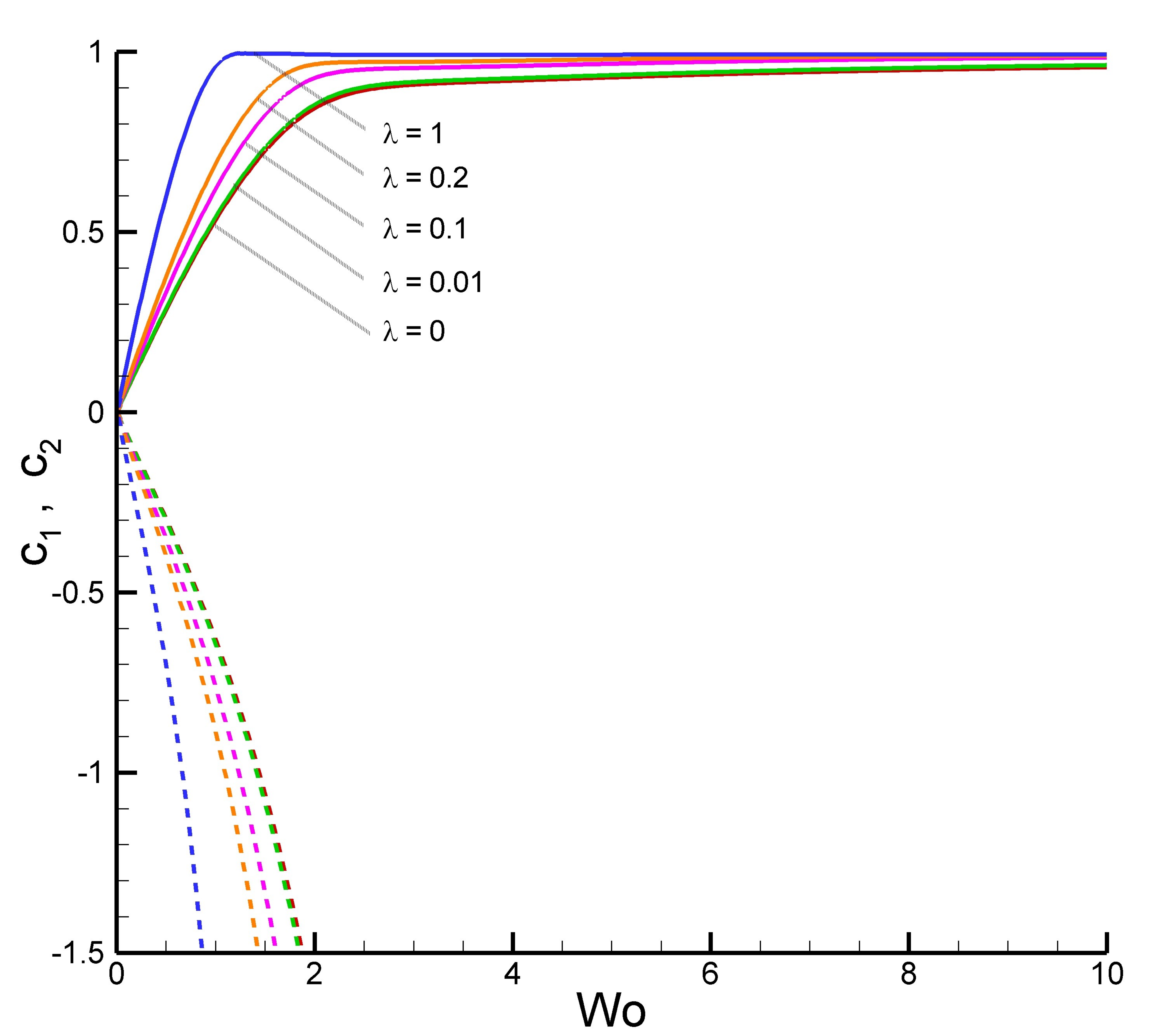}
\caption{Variation of the dispersion($c_1$) and attenuation($c_2$) coefficients with the frequency parameter Wo for various normalized slip ratios. Solid and dashed lines show $c_1/c_0$, and $c_2/c_0$, respectively. As the frequency parameter increases both attenuation and dispersion effects vanish with $c_2/c_0 \rightarrow - \infty$ and $c_1/c_0 \rightarrow 1$.}
\label{fig:c1c2}
\end{figure}

\begin{figure}
\centering
\includegraphics[width=0.75\textwidth]{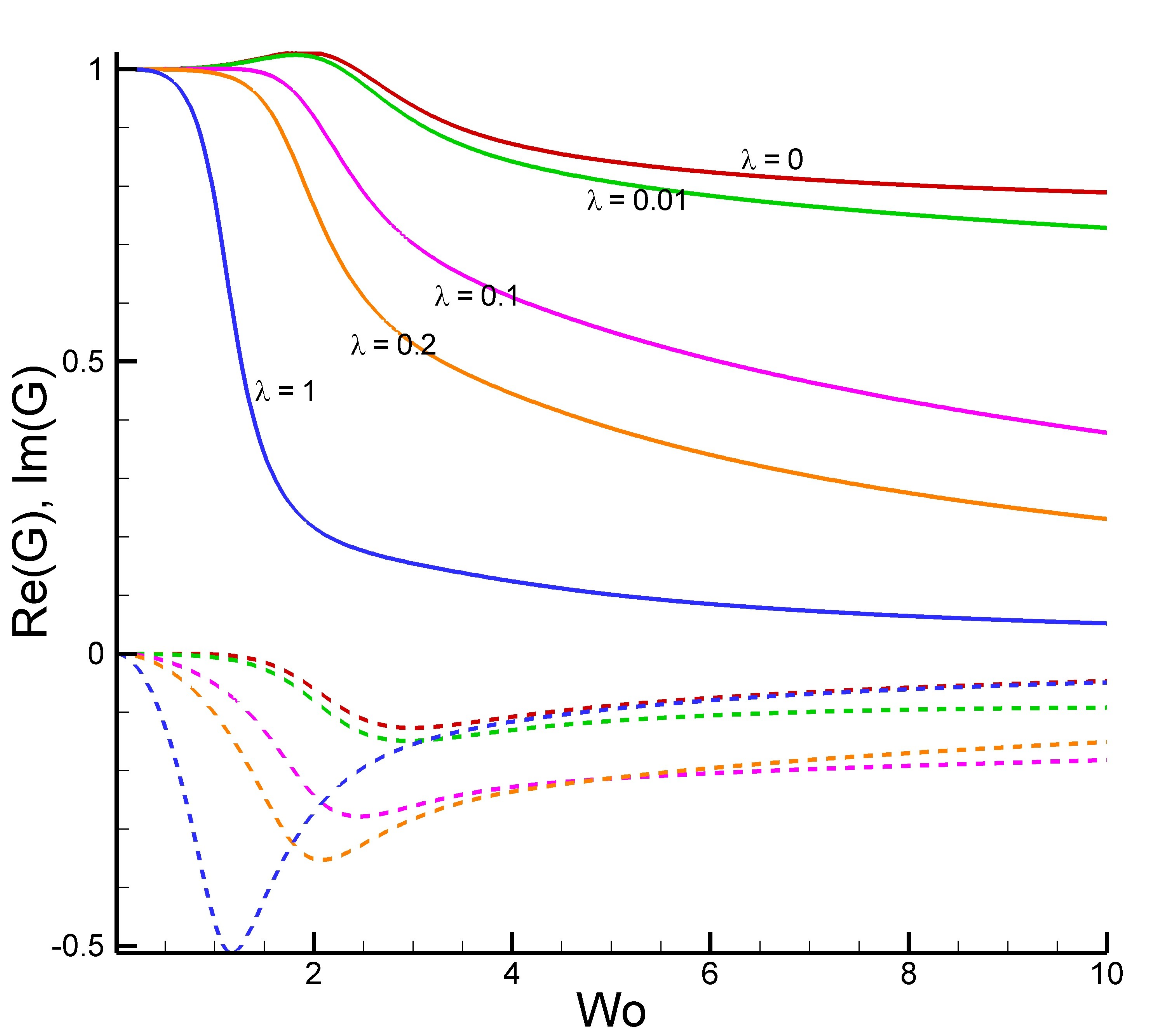}
\caption{Variation of the real (solid lines) and imaginary (dashed lines) parts of the elasticity factor $G$ for different values of $\lambda$.}
\label{fig:G}
\end{figure}

Since the governing equations are linear we seek a solution for oscillatory flow having the angular frequency $\omega$ of the form [Zamir, 2000]
\begin{eqnarray}\label{eq:system}
p(r,z,t)=P(r)e^{i\omega(t-z/c)}\\
u(r,z,t)=U(r)e^{i\omega(t-z/c)}\\
v(r,z,t)=V(r)e^{i\omega(t-z/c)}
\end{eqnarray}
where $c$, the wave propagation speed, is a complex variable. This assumption is limited to cases of harmonic oscillatory motion. It can, however, be generalized in terms of a Fourier series expansion of different harmonics since our equations are linear. If we define the following transformation
\begin{equation}
\zeta = \Lambda \frac{r}{R}, \quad \Lambda = (\frac{i-1}{\sqrt{2}}) \mbox{Wo}, \quad \mbox{Wo}=\sqrt{\frac{\rho \omega}{\mu}}R
\end{equation}
and substitute it into the governing equations we obtain the following ordinary differential equation for continuity
\begin{equation}
V^{\prime} + \frac{V}{\zeta} - \frac{i \omega R}{c \Lambda}U = 0
\end{equation}
and the following Bessel equation for axial momentum
\begin{equation}
\zeta^2 U^{\prime\prime} + \zeta U^{\prime} + \zeta^2 U = \frac{P\zeta^2}{\rho c}
\end{equation}
which has a general homogenous solution of the form
\begin{equation}
U^h(\zeta) = a_1 J_0(\zeta) + a_2 Y_0(\zeta)
\end{equation}
where $a_1$, and $a_2$ are integration constants and $J_0$, and $Y_0$ are the zeroth order Bessel functions of first and second kind [Yousif and Melka, 1997]. Since $U$ is bounded at $\zeta=0$, we see that $a_2=0$. The radial momentum equation becomes
\begin{equation}
\zeta^2 V^{\prime\prime} + \zeta V^{\prime} + (\zeta^2-1) V = \frac{i\Lambda P^{\prime} \zeta^2}{\rho R \omega}
\end{equation}
which has a homogenous solution of the form
\begin{equation}
V^h(\zeta) = b_1 J_1(\zeta) + b_2 Y_1(\zeta)
\end{equation}
where $b_1$, and $b_2$ are integration constants and $J_1$, and $Y_1$ are the first order Bessel functions of the first and second kind. Similarly, $V$ is bounded at $\zeta=0$, and $b_2=0$. General solutions satisfying the governing differential equations then become
\begin{equation}
P(r) = B  J_0(\frac{i\omega r}{c})
\end{equation}
\begin{equation}
U(r) = A J_0(\Lambda \frac{r}{R}) + \frac{i B }{c(i \rho - \mu \omega/c )}J_0(\frac{i\omega r}{c})
\end{equation}
\begin{equation}
V(r) =  A \frac{i \omega R}{c \Lambda}J_1(\Lambda \frac{r}{R}) + \frac{i B }{c(i \rho - \mu \omega/c )}J_1(\frac{i\omega r}{c}).
\end{equation}
If we use the following simplifications: $J_0(\frac{i\omega r}{c})\approx1$, $J_1(\frac{i\omega r}{c})\approx\frac{i\omega r}{2c}$, and $\rho \gg \mu \omega/c$, which are valid under assumption (i), the solution of the system of differential equations becomes
\begin{equation}\label{eq:ode1}
P(r) = B
\end{equation}
\begin{equation}\label{eq:ode2}
U(r) = A J_0(\Lambda \frac{r}{R}) + \frac{B}{\rho c}
\end{equation}
\begin{equation}\label{eq:ode3}
V(r) = A \frac{i \omega R}{c \Lambda}J_1(\Lambda \frac{r}{R}) + B \frac{i\omega}{2 \rho c^2}r.
\end{equation}
Here, $A$ and $B$, the integration constants, and $c$, the wave propagation speed, remain to be determined. It is reasonable to assume that the radial and axial oscillatory movements of the tube have the same frequency as that prevailing in the flow field, so that
\begin{equation}
\xi(z,t) = C e^{i\omega(t-z/c)}
\end{equation}
\begin{equation}
\eta(z,t) = D e^{i\omega(t-z/c)}.
\end{equation}
This does not, however, imply that the wall motion in phase with oscillatory motion of the fluid [Zamir, 2000]. Therefore, we have additional two constant $C$ and $D$ that need to be determined. If we use the the shell equations along with the matching conditions given in the previous section we get a set of four homogenous linear equations for $A$, $B$, $C$, and $D$ with \emph{rank three} which can be written as
\begin{equation}
\left[ \begin{array}{cccc}
\frac{\mu \Lambda J_1(\Lambda)}{\rho_w h R}& 0 & \omega^2(1- \frac{E h}{\rho_w (1-\nu^2) c^2}) & \frac{-i \omega \nu E}{R \rho_w (1-\nu^2) c} \\
0   & \frac{1}{h}  & \frac{i \omega \nu E}{R (1-\nu^2) c}  & \frac{-E}{R^2 (1-\nu^2)}\\
J_0(\Lambda) - \frac{\ell}{R}\Lambda  J_1(\Lambda)     & \frac{1}{\rho c} & - i \omega  & 0 \\
\frac{i \omega R J_1(\Lambda) }{\Lambda c}  & \frac{i \omega R}{2 \rho c^2} & 0 & -i\omega
\end{array} \right]  \left\{ \begin{array}{c} A \\ B \\ C \\ D \end{array} \right\} =  0.
\label{eq:1}
\end{equation}
This matrix equation has a nontrivial solution if the determinant of the coefficient matrix is zero, and gives an equation for the wave propagation speed. After some algebra the quadratic dispersion relationship becomes
\begin{equation}
\label{eq:dis}
\alpha \theta^2 + \beta \theta + \gamma = 0
\end{equation}
where
\begin{equation}
\alpha = (1-\nu^2)(1-g-\chi)
\end{equation}
\begin{equation}
\beta = g(2\nu-\frac{1}{2}) - \kappa(1-g-\chi) - 2(1-\chi)
\end{equation}
\begin{equation}
\gamma = g+2\kappa(1-\chi)
\end{equation}
where
\begin{equation}
\theta = \frac{Eh}{(1-\nu^2)\rho R c^2}, \kappa=\frac{\rho_w h}{\rho R}, \chi = \Lambda\frac{\ell_s}{R}\frac{J_1(\Lambda)}{J_0(\Lambda)}, g=\frac{1}{\Lambda}\frac{J_1(\Lambda)}{J_0(\Lambda)}.
\end{equation}
Using the positive solution of Eq.\ref{eq:dis}
\begin{equation}
\theta = \frac{-\beta + \sqrt{\beta^2 - 4\alpha\gamma}}{2\alpha}
\end{equation}
the wave propagation speed becomes
\begin{equation}
\frac{c}{c_0}=\sqrt{\frac{2}{(1-\nu^2)\theta}}
\end{equation}
where we use the Moens-Korteweg inviscid wave propagation speed, $c_0$, for normalization,  which is defined as
\begin{equation}\label{eq:c0}
c_0=\frac{Eh}{2\rho R}.
\end{equation}
Here, the parameter $\theta$, and the wave propagation speed, $c$, are functions of the fluid properties and tube structural properties, geometry and slip length. We define the slip ratio as $\lambda=\frac{\ell_s}{R}$ and its influence on the flow characteristics is presented in this study. The variation of the real and imaginary parts of the wave propagation speed $c$, normalized in terms of the inviscid wave speed, $c_0$, with frequency parameter, Wo, for various slip ratios is shown in Fig.\ref{fig:c}. Here, $\lambda=0$ corresponds to the classical Womersley solution for elastic tubes. As $\lambda$ increases, both the real and imaginary parts of the wave propagation speed increase for certain range of Wo numbers. As the frequency increases, the imaginary part of $c$ vanishes while the real part becomes the same as inviscid wave propagation speed.
\begin{figure}
\centering
\includegraphics[width=0.75\textwidth]{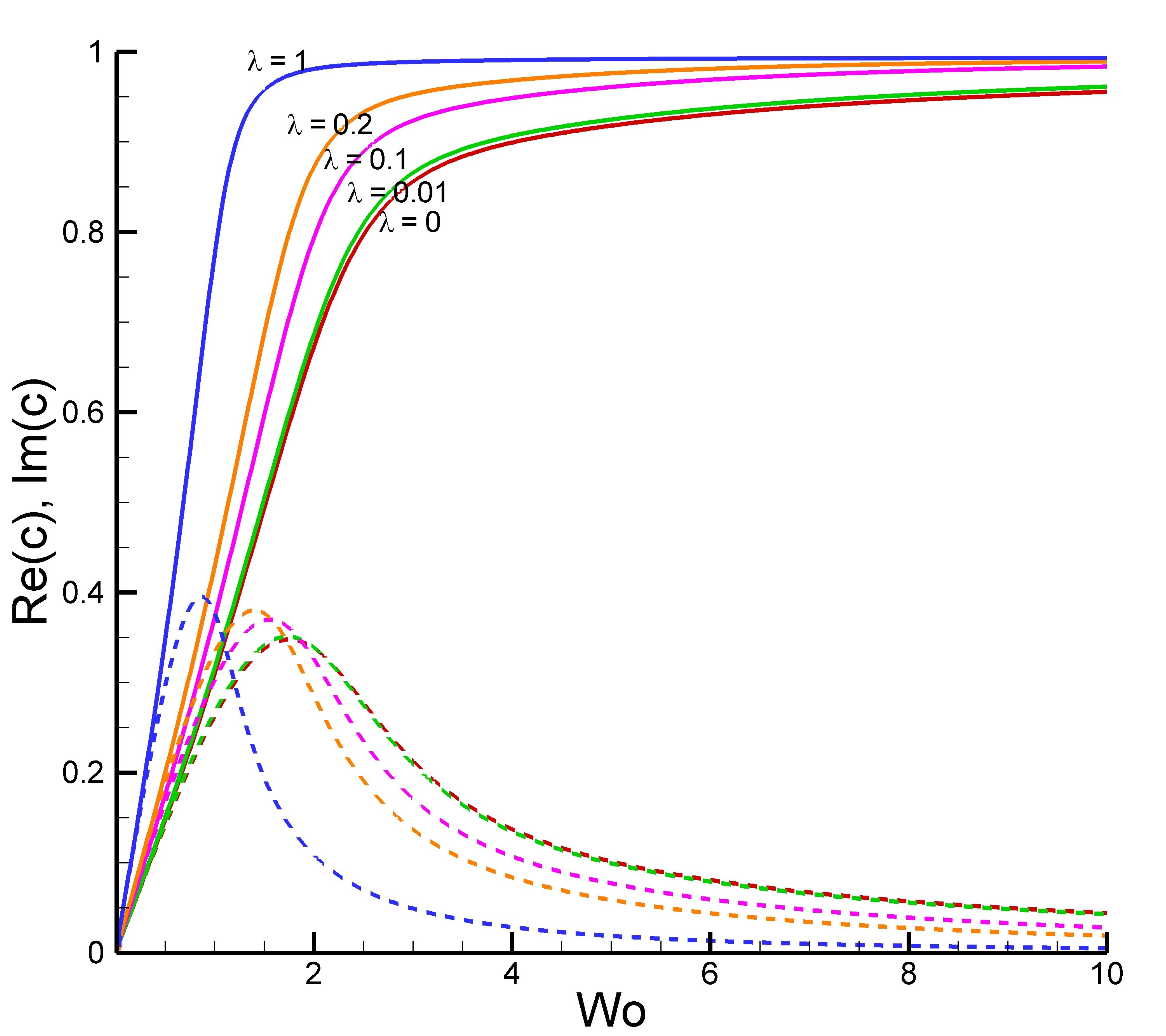}
\caption{Variation of the real (solid lines) and imaginary (dashed lines) parts of the wave speed $c$ with frequency parameter, Wo. The wave speed is normalized by the inviscid wave speed $c_0$  for various normalized slip ratios $\lambda=\frac{\ell_s}{R}$. }
\label{fig:c}
\end{figure}
The effects of the wave propagation speed can be seen in the solution via the term $e^{i\omega(t-z/c)}$. If we define
\begin{equation}
\frac{1}{c}=\frac{1}{c_1}+i\frac{1}{c_2}
\end{equation}
where $c_1$ and $c_2$ are dispersion and attenuation coefficients, then the term becomes $e^{\omega z/c_2} e^{i\omega(t-z/c_1)}$. Variations of the dispersion and attenuation coefficients with frequency parameter are shown in Fig.\ref{fig:c1c2} for different slip ratios. As the frequency parameter increases $c_2/c_0 \rightarrow - \infty$ and $c_1/c_0 \rightarrow 1$, and both attenuation and dispersion effects vanish earlier with increasing slip ratios $\lambda$.
In addition to the wave propagation speed, as a consequence of the \emph{rank three} linear system of the four equations, the solutions for the coefficients $A$, $C$, and $D$ are expressed in terms of $B$ as
\begin{equation}\label{eq:coefA}
A = (\frac{B}{\rho c J_0(\Lambda)})(\frac{2 + \theta(2\nu-1)}{\theta(g-2\nu(1-\chi))})
\end{equation}
\begin{equation}\label{eq:coefC}
C = (\frac{iB}{\rho c \omega})(\frac{\theta(1-g-\chi)-2(1-\chi)}{\theta(g-2\nu(1-\chi))})
\end{equation}
\begin{equation}\label{eq:coefD}
D = (\frac{BR}{\rho c^2})(\frac{g-\theta\nu(1-g-\chi)}{\theta(g-2\nu(1-\chi))}).
\end{equation}
Considering the solution for pressure
\begin{equation}
p(r,z,t) = B e^{i\omega(t-z/c)}
\end{equation}
it can be seen that the pressure gradient becomes
\begin{equation}
\frac{\partial p}{\partial z} = -\frac{iB\omega}{c} e^{i\omega(t-z/c)}.
\end{equation}
Prescribing the amplitude of the pressure gradient completes the analytical solutions derived here and closes the system of equations. To provide a reference solution, we choose the amplitude of pressure gradient to be the steady-state Poiseuille flow pressure gradient, $\Phi$, setting $B=\frac{i\Phi c}{\omega}$. We normalize our solutions with respect to the steady-state Poiseuille flow solution, which is given as
\begin{equation}
u_s = -\frac{R^2}{4\mu}\Phi[1-(\frac{r}{R})^2]
\end{equation}
with the corresponding flow rate and shear stress
\begin{equation}
q_s = -\frac{\pi R^4}{8\mu}\Phi, \quad (\tau_w)_s = -\frac{R}{2}\Phi
\end{equation}
and we use the following centerline ($r=0$) velocity to normalize the equations
\begin{equation}
\bar{u}_s = -\frac{R^2}{4\mu}\Phi.
\end{equation}
Interestingly, the solution of the steady flow with slip boundary condition results in following field
\begin{equation}\label{eq:sslip}
u^{slip}_s = \bar{u}_s [1-(\frac{r}{R})^2 + \frac{2\ell_s}{R}]
\end{equation}
\begin{equation}
q^{slip}_s = -\frac{\pi R^4}{8\mu}\Phi[1+\frac{4\ell_s}{R}]
\end{equation}
\begin{equation}
(\tau_{w})_{s}^{slip} = (\tau_{w})_s
\end{equation}
where Navier slip boundary condition have been implemented. The term $\frac{2\ell_s}{R}$ in Eq.~\ref{eq:sslip} represent an additional velocity associated with the Navier slip boundary conditions. For finite $\ell_s$, as the radius of tube decreases the importance of this term increases. Consequently, as discussed by Wu {\it et al.} [2008], the implementation of slip boundary conditions adds a rigid body translation term to the axial velocity. The slip solution becomes a superposition of the no-slip solution and rigid body translation. Since rigid body translation does not lead to any change in the material deformation the shear stress is not affected. When slip boundary conditions are introduced to transient oscillatory flow, the effect on the solution is not simply the addition of a rigid body translation term, and nonlinear behavior results. Wu {\it et al.} [2008] examined this subject by considering oscillatory flows through rigid tubes. In this study, we extended his study to flow through the elastic tubes, and by considering the axial and radial momentum equations and the cylindrical shell equations. If we substitute Eqs.~\ref{eq:ode1}-\ref{eq:ode3} and the corresponding coefficients given in Eqs.~\ref{eq:coefA}-\ref{eq:coefD} into the Eq.~\ref{eq:system}, the following normalized solutions are obtained
\begin{equation}\label{eq:GEs}
\frac{u(r,z,t)}{\bar{u}_s} =  -\frac{4}{\Lambda^2}[1-G\frac{J_0(\Lambda \frac{r}{R})}{J_0(\Lambda)}]e^{i\omega(t-z/c)}
\end{equation}
\begin{equation}
\frac{q(z,t)}{q_s} =  -\frac{8}{\Lambda^2}[1-Gg]e^{i\omega(t-z/c)}
\end{equation}
\begin{equation}
\frac{\tau_w(z,t)}{(\tau_w)_s} =  [Gg]e^{i\omega(t-z/c)}
\end{equation}
\begin{equation}
\frac{v(r,z,t)}{\bar{u}_s}  =  \frac{2 \omega R}{i \Lambda^2 c}[\frac{r}{R}-Gg\frac{J_1(\Lambda \frac{r}{R})}{J_1(\Lambda)}]e^{i\omega(t-z/c)}
\end{equation}
\begin{equation}\label{eq:xi}
\frac{\xi(z,t)}{R}  = \frac{C}{R} e^{i\omega(t-z/c)}
\end{equation}
\begin{equation}\label{eq:eta}
\frac{\eta(z,t)}{R}  = \frac{D}{R} e^{i\omega(t-z/c)}
\end{equation}
where $G$, the elasticity factor, an indicator of the difference between elastic and rigid tubes, is given as
\begin{equation}
G = -\frac{\rho c J_0(\Lambda)}{B} = \frac{2+\theta(2\nu-1)}{\theta(2\nu(1-\chi)-g)}
\end{equation}
where $\theta$ comes from dispersion relationship, Eq.~\ref{eq:dis}, and is a function of the slip parameter $\chi$. It can clearly be seen that the effect of slip boundary conditions is nonlinear in this case. The variation of the real and imaginary parts of the elasticity factor with Wo are illustrated in Fig.~\ref{fig:G} for different $\lambda$ values. As a special case, when $\chi=0$, corresponding to no-slip boundary conditions, the solution reduces to the classical Womersley's solution for elastic tubes. Additionally, for $G=1$ and $c\rightarrow\infty$ the solution reduces to Womersley's solution for the rigid case.

\section{Results}
\label{sec:Results}
Solutions of the governing equations for oscillatory flow through the elastic tubes were derived using Navier slip boundary conditions in Section 3, and the corresponding analytical solutions  given in Eqs.~\ref{eq:GEs}-\ref{eq:eta}. In the following results, the constant values $c_0=10 m/s$ and $\kappa=0.1$ are used. The parametric variables are Womersley number, Wo, and the slip ratio, $\lambda$.
\begin{figure}
\centering
\mbox{\subfigure{\includegraphics[width=0.5\textwidth]{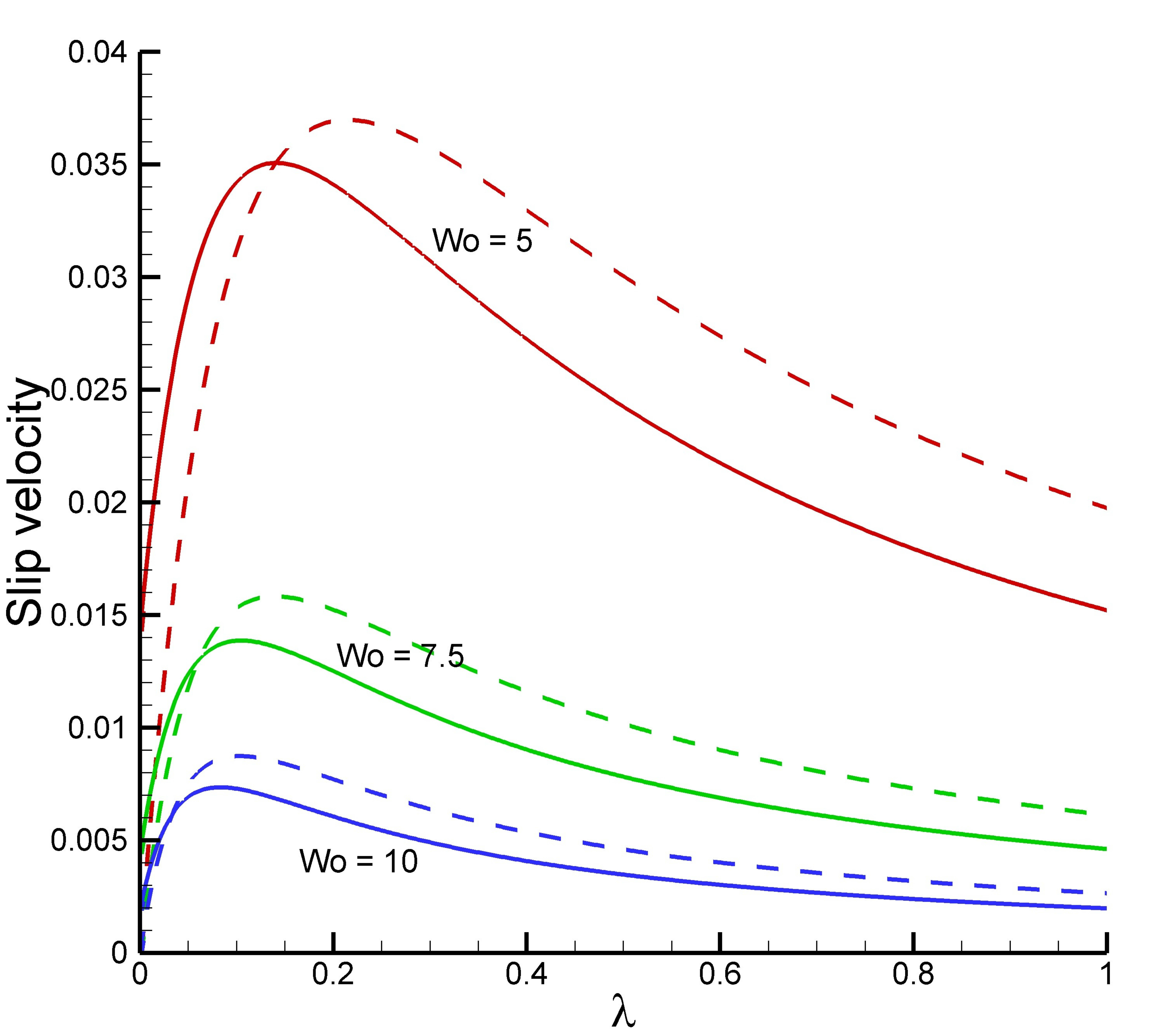}}
\subfigure{\includegraphics[width=0.5\textwidth]{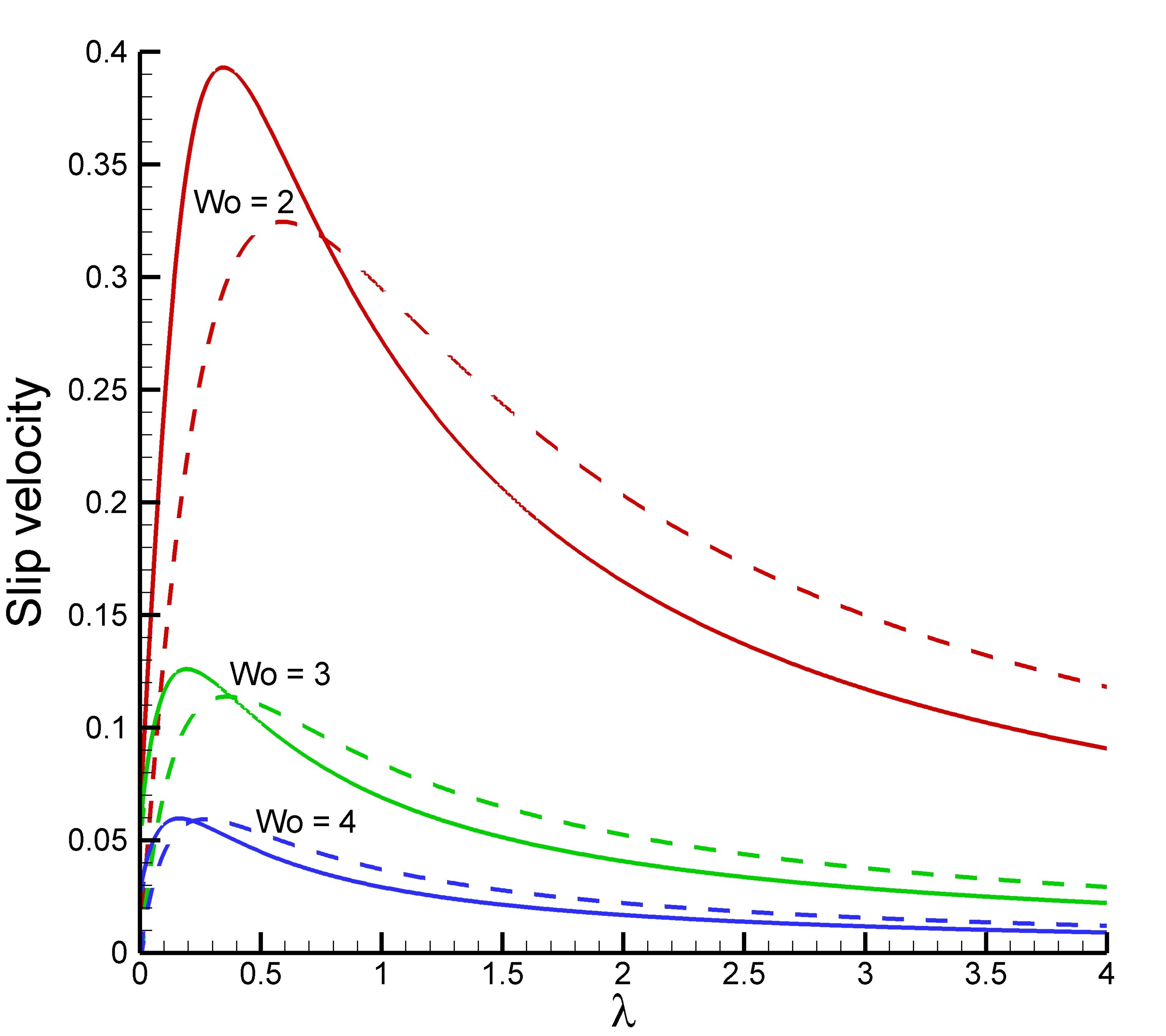} } }
\\
\mbox{\subfigure{\includegraphics[width=0.5\textwidth]{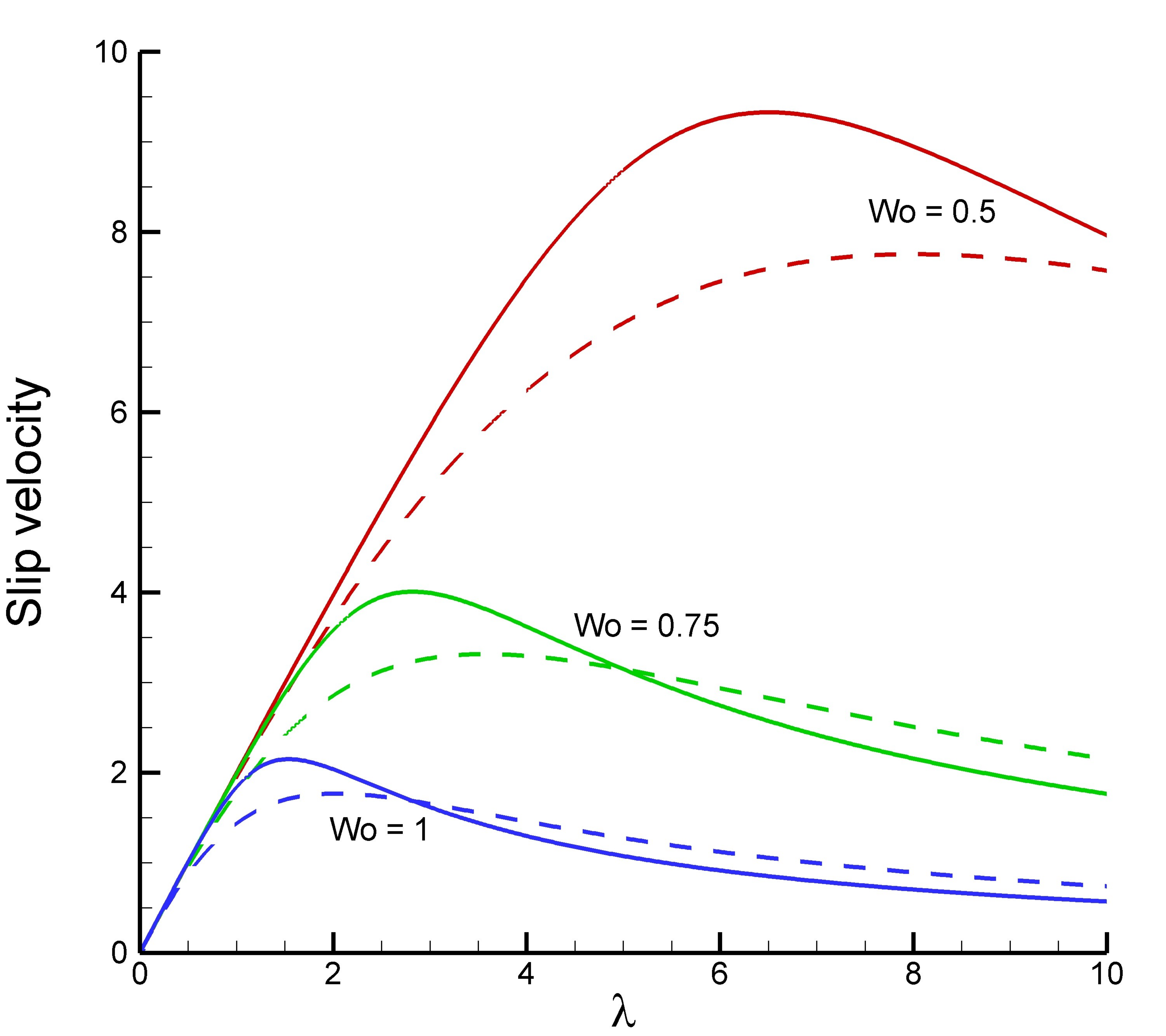}}
\subfigure{\includegraphics[width=0.5\textwidth]{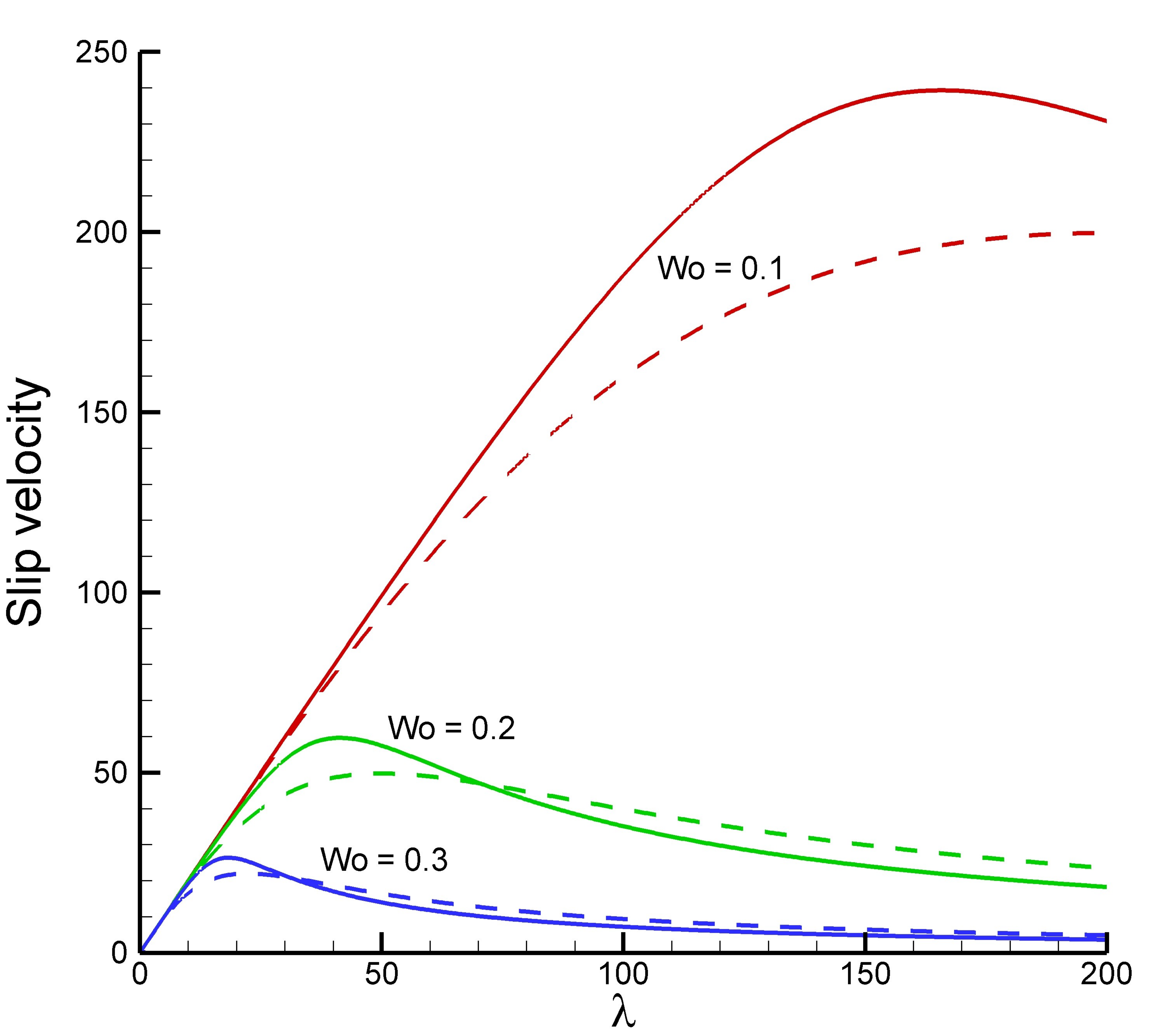}}}
\caption{Influence of the slip ratio $\lambda$ on the normalized slip velocity at the wall. Solid lines show the elastic wall case and dashed lines show the rigid case: (a) $\mbox{Wo}=10$, $\mbox{Wo}=7.5$, $\mbox{Wo}=5$; (b) $\mbox{Wo}=4$, $\mbox{Wo}=3$, $\mbox{Wo}=2$; (c) $\mbox{Wo}=1$, $\mbox{Wo}=0.75$, $\mbox{Wo}=0.5$; (d) $\mbox{Wo}=0.3$, $\mbox{Wo}=0.2$, $\mbox{Wo}=0.1$.}
\label{fig:slip}
\end{figure}

The influence of the slip length on the slip velocity at the surface is shown in Fig.~\ref{fig:slip} for various Womersley numbers. Results for the rigid wall case are also included. In order to make the presentation clear we present four different figures with different $\lambda$ scales. As can be seen in the figures, the slip velocities at the wall increase rapidly with increasing slip length and then, after a certain critical value of the slip length, reduce slowly. Comparing the elastic and rigid tube results, one can see that the maximum slip velocity can be obtained at a smaller slip ratio for the elastic tubes. It can also easily be seen that for large Wo there is a contribution to the slip velocity, which is due to axial deformation. For small Wo we extend the range of $\lambda$ to be able to see the full behavior of the slip velocity. This can be related to nanoscale experimental observations that the slip length can be much bigger than the tube radius at the nanoscale. Our results demonstrate the same conclusions. We note, however, that these large values of $\lambda$ may have physical meaning only for liquid flows at the nanoscale, which correspond to nearly frictionless flows. Majumder {\it et al.} [2005] measured the slip length of a carbon nanotube with a diameter of $7$nm to be about $3-70\mu$m for different liquid flows (water, ethanol, propanol, hexane, and decane). In the rest of the paper we restrict the results to those for small $\lambda$ to make the analysis applicable for a wide range of systems.
\begin{figure}
\centering
\includegraphics[width=0.75\textwidth]{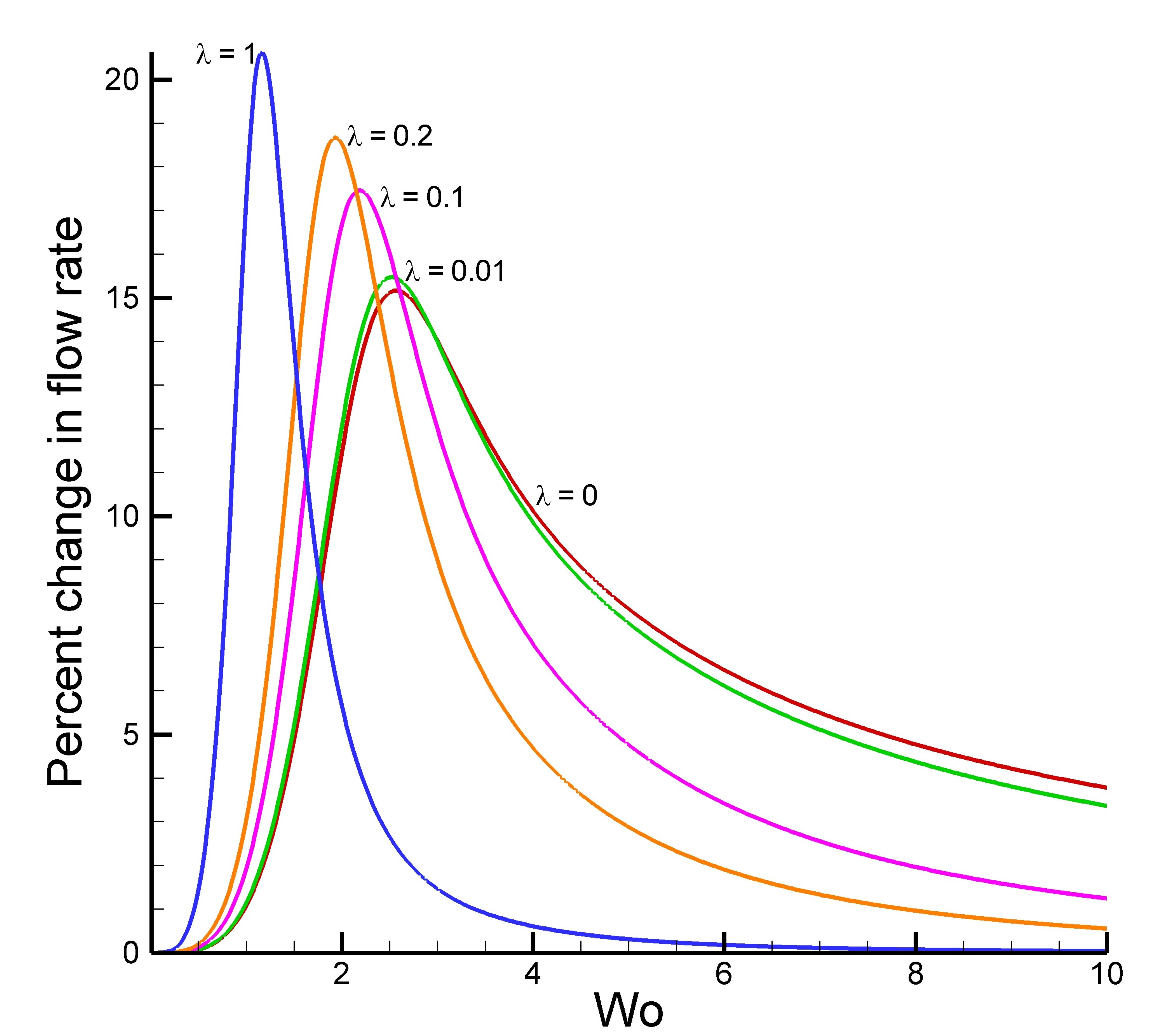}
\caption{Percent difference between the peak flow rate in an elastic tube and that in a rigid tube for various slip ratios.}
\label{fig:rate}
\end{figure}

The amplitude of the percent difference between the peak flow rate in an elastic tube and that in a rigid tube is shown in Fig.~\ref{fig:rate} for various slip ratios. By increasing the slip length at the wall the difference between the rigid and elastic tubes reaches its peak values at a smaller value of Wo. We can also interpret this as being for a fixed frequency and smaller size microtubes ($\mbox{Wo}=\sqrt{\mbox{Re} \mbox{St}}$, where Re and St are the Reynolds and Strouhal numbers). It is already established that the wall movements in an elastic tube make it easier for the flow to move through the tube than through a rigid tube [Zamir, 2000]. Here we show, in addition, that the the velocity slip at the surface also has the same behavior, with increasing sensitivity to Wo. Therefore, the driving frequency has even more of an effect in controlling the flows through smaller devices in pulsatile microfluidics.
\begin{figure}
\centering
\includegraphics[width=0.75\textwidth]{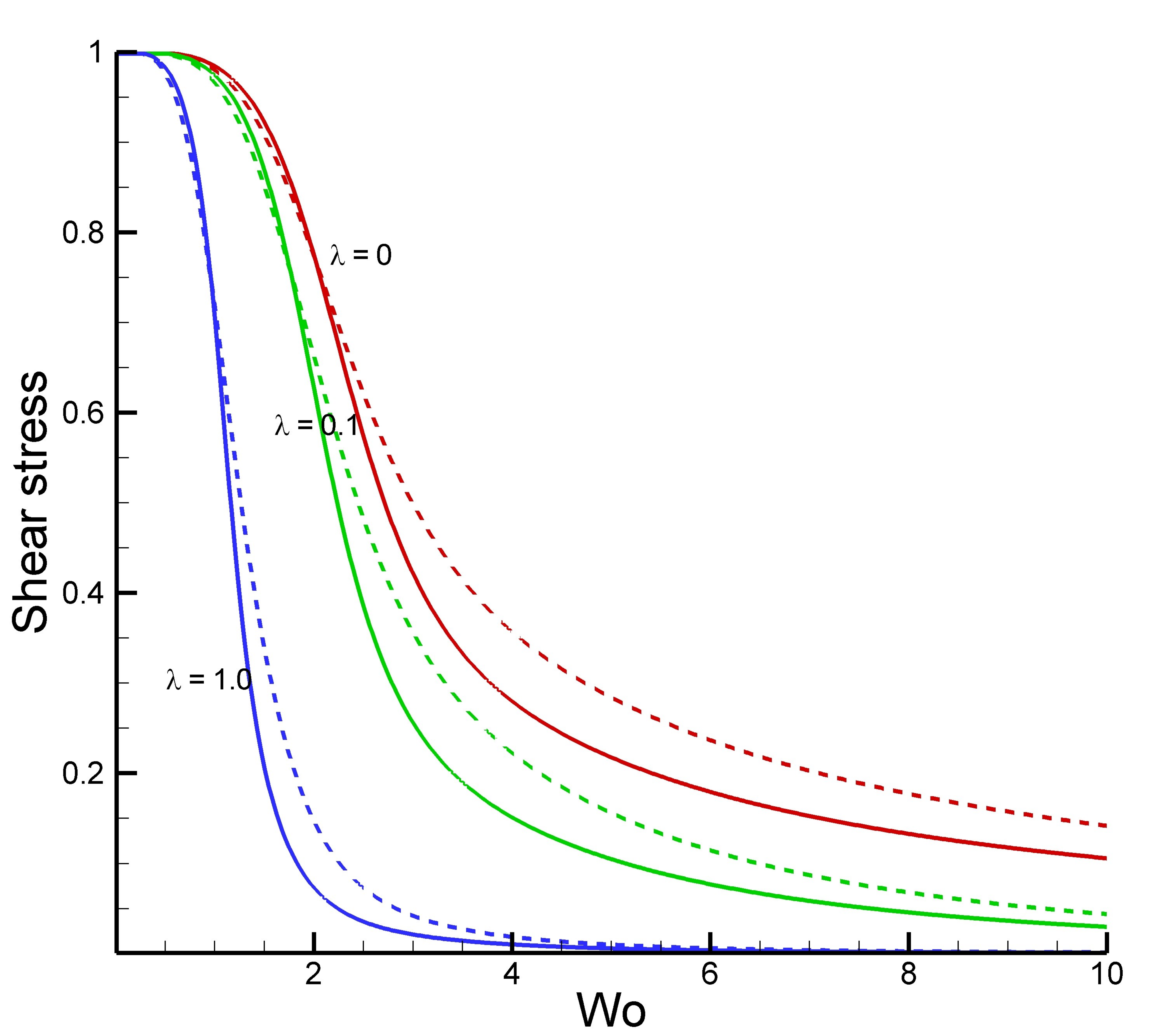}
\caption{Normalized wall shear stresses amplitudes. The solid and dashed lines show the elastic and rigid wall cases, respectively.}
\label{fig:shear}
\end{figure}

The normalized shear stress amplitudes at the wall are shown in Fig.~\ref{fig:shear} for both elastic and rigid walls. The solid and dashed lines show the slip ($\lambda=0.1$) and no-slip ($\lambda=0$) cases, respectively. It can be seen that the shear stress at the walls in elastic tubes is greater than that in rigid tubes for smaller Wo. There is always a critical Wo at which the rigid and elastic solutions intersect, and this critical Wo number decreases with increasing slip ratio. It can be also seen that the shear stress at the wall decreases with decreasing Wo and nearly coincides with the frictionless flow measurements at the smallest values.

\begin{figure}
\centering
\includegraphics[width=\textwidth]{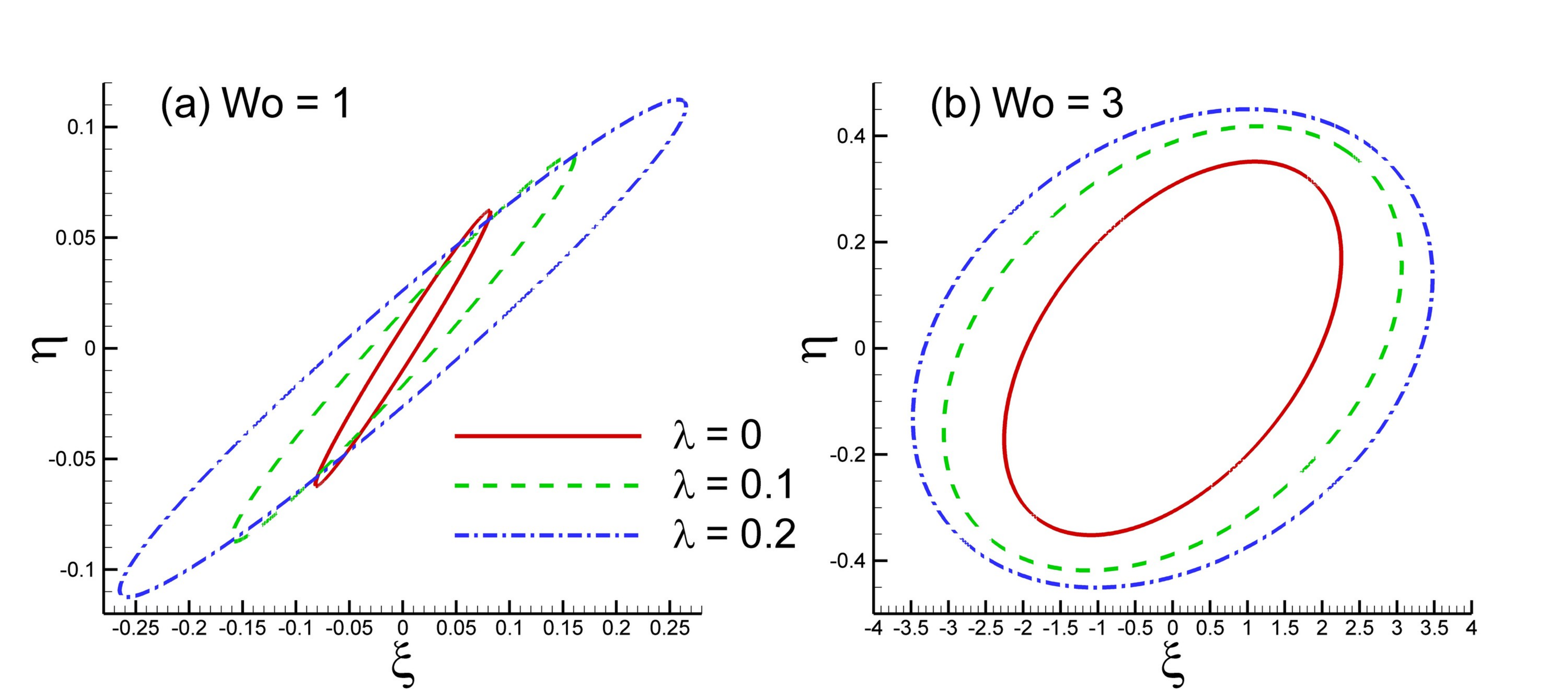}
\caption{Particle trajectories of the tube wall using both radial and axial displacements. These are given by Eqs.~\ref{eq:xi} and~\ref{eq:eta}, which are normalized here by $\frac{i \Phi }{\rho \omega c}$.}
\label{fig:eta}
\end{figure}

In the previous figures we analyzed the amplitude of the flow rate, the slip-velocity, and the shear stress at the walls, and compared the results for both elastic and rigid tubes with different slip ratios. For those cases, the exponential terms in the solutions took unit value (i.e., $z=0$, $t=0$). Now, we introduce the axial tube length, $z:[0,L_t]$, and the time with respect to one oscillatory period, $t:[0,T]$. We assume that the the tube length is much greater than the tube radius, and that the wavelength, $L_w$, is also much greater than the tube radius ($L_t >> R$). Therefore, in the limit of the long-wavelength assumption, the wavelength $L_w$ can be of the same order or greater than the tube length $L_t$. The wavelength is defined as $L_w = cT =2 \pi c/\omega$.

The particle trajectories of the tube wall using radial and axial displacements given in Eqs.~\ref{eq:xi} and~\ref{eq:eta} are shown in Fig.~\ref{fig:eta} for $\mbox{Wo}=1$ and $\mbox{Wo}=3$. The displacements are normalized by $\frac{i \Phi }{\rho \omega c}$ (i.e., the axes in the figure are $\xi /[iR\Phi/(\rho \omega c)]$, $\eta /[iR\Phi/(\rho \omega c)]$). As shown in the figure, the trajectories are elliptic and their areas increase for increasing slip ratios $\lambda$. The solid line shows the no-slip ($\lambda=0$) case, and the dashed and dash-dotted lines show the slip cases for $\lambda=0.1$ and $\lambda=0.2$. It can be seen that the amplitudes of the displacements increase with increasing slip ratio. Additionally, the aspect ratios of the elliptical trajectories become smaller with increasing Womersley numbers. The orientation of the elliptical axes is also changing by increasing the slip ratio. For higher slip ratios (i.e., smaller scales), we can see that the slope of the longitudinal axis of the elliptical trajectories gradually decreases.
\begin{figure}
\centering
\includegraphics[width=\textwidth]{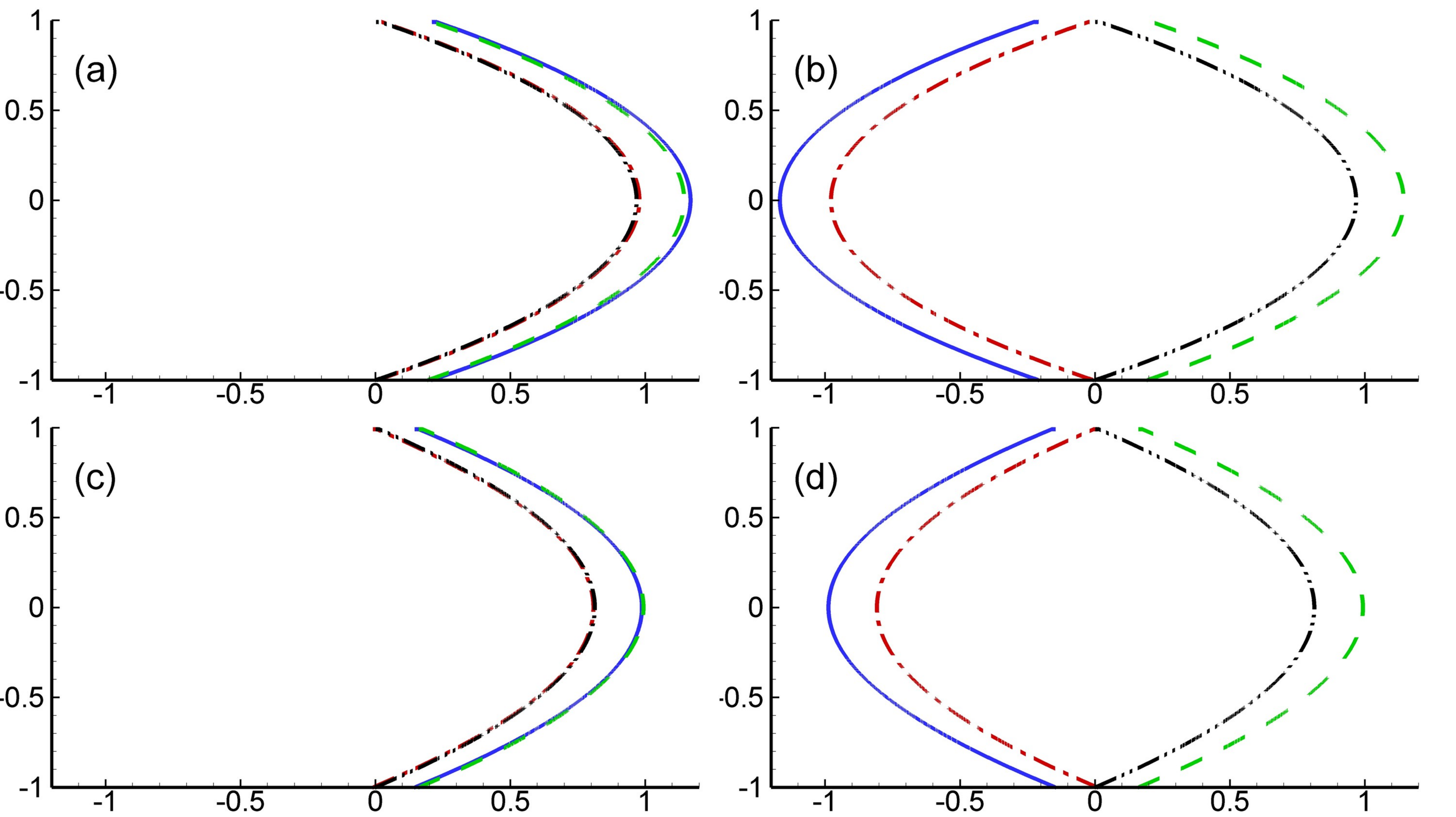}
\caption{Normalized velocity profiles for $\mbox{Wo}=1$ and $L_t/L_w=1$ at (a) $t/T=0$, $z/L_t = 0$; (b) $t/T=0$, $z/L_t = 1/2$; (c) $t/T=1/4$, $z/L_t = 0$; and (d) $t/T=1/4$, $z/L_t = 1/2$. Red dash-dotted line: no-slip elastic wall, black dash-double dotted line: no-slip rigid wall, blue solid line: slip elastic wall, green dashed line: slip rigid wall. }
\label{fig:a}
\end{figure}
\begin{figure}
\centering
\includegraphics[width=\textwidth]{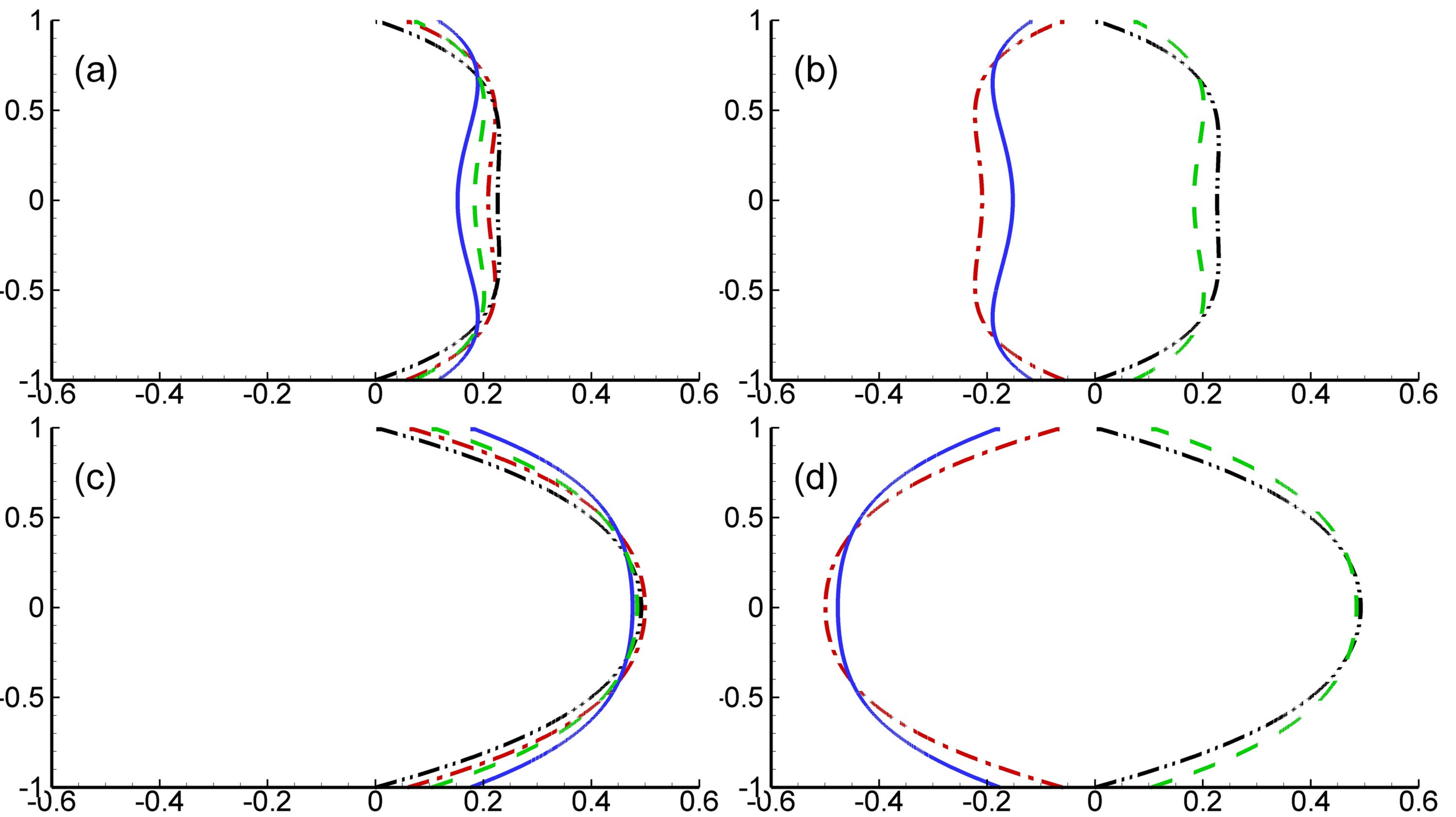}
\caption{Normalized velocity profiles for $\mbox{Wo}=3$ and $L_t/L_w=1$ at (a) $t/T=0$, $z/L_t = 0$; (b) $t/T=0$, $z/L_t = 1/2$; (c) $t/T=1/4$, $z/L_t = 0$; and (d) $t/T=1/4$, $z/L_t = 1/2$. Red dash-dotted line: no-slip elastic wall, black dash-double dotted line: no-slip rigid wall, blue solid line: slip elastic wall, green dashed line: slip rigid wall. }
\label{fig:b}
\end{figure}
\begin{figure}
\centering
\includegraphics[width=\textwidth]{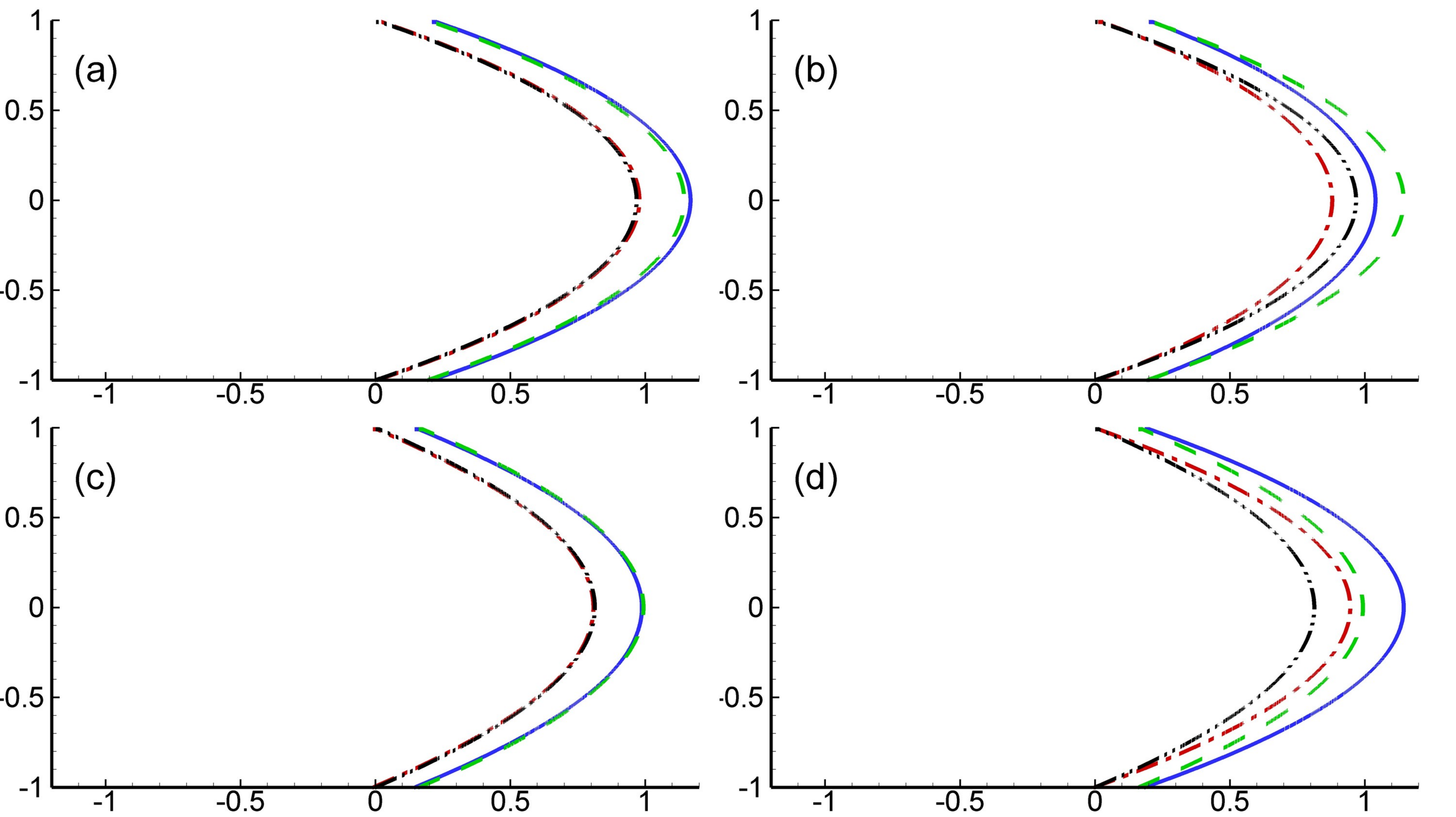}
\caption{Normalized velocity profiles for $\mbox{Wo}=1$ and $L_t/L_w=0.1$ at (a) $t/T=0$, $z/L_t = 0$; (b) $t/T=0$, $z/L_t = 1/2$; (c) $t/T=1/4$, $z/L_t = 0$; and (d) $t/T=1/4$, $z/L_t = 1/2$. Red dash-dotted line: no-slip elastic wall, black dash-double dotted line: no-slip rigid wall, blue solid line: slip elastic wall, green dashed line: slip rigid wall. }
\label{fig:c}
\end{figure}
\begin{figure}
\centering
\includegraphics[width=\textwidth]{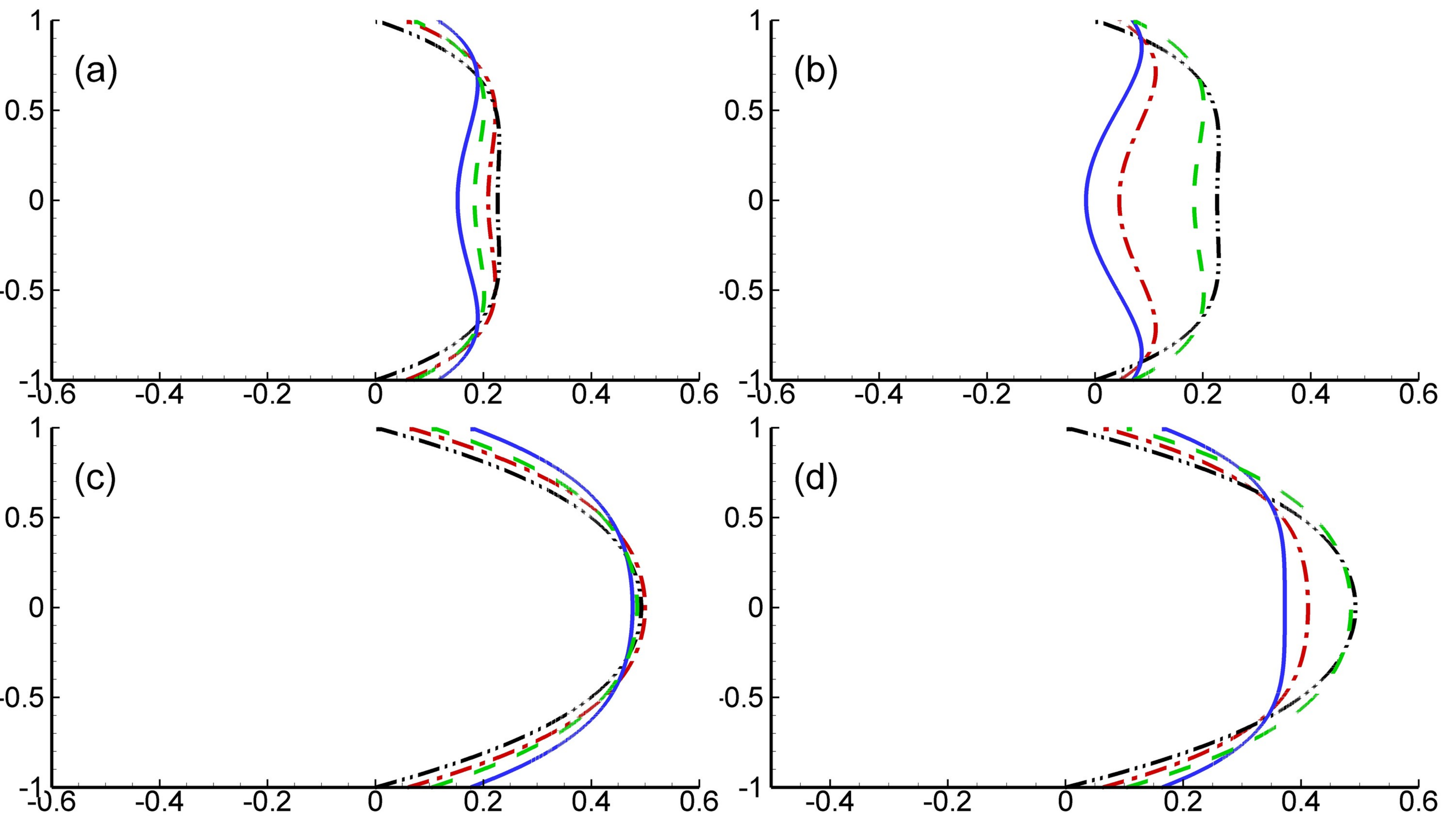}
\caption{Normalized velocity profiles for $\mbox{Wo}=3$ and $L_t/L_w=0.1$ at (a) $t/T=0$, $z/L_t = 0$; (b) $t/T=0$, $z/L_t = 1/2$; (c) $t/T=1/4$, $z/L_t = 0$; and (d) $t/T=1/4$, $z/L_t = 1/2$. Red dash-dotted line: no-slip elastic wall, black dash-double dotted line: no-slip rigid wall, blue solid line: slip elastic wall, green dashed line: slip rigid wall. }
\label{fig:d}
\end{figure}

The following results for the axial velocity are presented to characterize the flow properties in elastic microtubes. The normalized axial velocity profiles according to Eq.~\ref{eq:GEs} for different frequency parameters and tube lengths are shown in Figs.~\ref{fig:a}-\ref{fig:d}. We include the solutions corresponding to the rigid case as well. We also compare with the no-slip and Navier slip boundary conditions cases. In these figures, the slip ratio is chosen to be $\lambda=0.1$, so that the slip length at the wall is ten times smaller than the tube radius. In each subfigure the line styles and colors indicate the following cases: the red dash-dotted line: no-slip elastic wall, the black dash-doubledotted line: no-slip rigid wall, the blue solid line: slip elastic wall, and the green dashed line: slip rigid wall. The profiles are presented at two discrete axial locations and for two different phases.

In Fig.~\ref{fig:a} and Fig.~\ref{fig:b}, $\mbox{Wo}=1$ and $\mbox{Wo}=3$, respectively, for the tube length to wavelength ratio $\frac{L_t}{L_w}=1$. When Wo is lower as in Fig.~\ref{fig:a}, the velocity profiles become close to parabolic profiles. As discussed earlier, the presence of slip at the solid surface has only translational effects for constant pressure driven flows. When we consider pulsatile flows, however, the slip length has a nonlinear effect. Nonlinear effects of the slip length are obvious for the elastic wall case as well, as is demonstrated in the figures. We see that the solutions for a rigid tube do not have any wave motion, so the velocity profiles at different sections are always the same at the same time (the wave propagation speed, $c\rightarrow\infty$). We can see the wave motion in the elastic tubes, however, due to the finite wave propagation speed; this phenomena can easily be seen because of the comparable tube lengths and wavelengths. The corresponding velocity profiles for the same tube length ratio but moderate frequency parameter $\mbox{Wo}=3$ are shown in Fig.~\ref{fig:b}. Since the frequency parameter is moderate we see that the velocity profiles are not parabolic and the amplitude is reduced at the centerline. Furthermore, if we compare the slip and no-slip solutions, we see that this reduced centerline velocity effect increases with slip velocity at the wall. This is due to fluid inertial effects. For larger Wo the flow has inertia acting in the opposite direction when the pressure gradient is reversed. Therefore, it will take some time before the pressure gradient can reverse the direction of the inertia. This time delay introduces a phase shift between the fluid motion and the pressure gradient. In a confined domain such as a tube, smaller velocities with correspondingly lower inertia towards the wall have smaller phase shifts than the fluid velocities in the center of the tube.

Fig.~\ref{fig:c} and Fig.~\ref{fig:d} show the velocity profile for $\mbox{Wo}=1$ and $\mbox{Wo}=3$ for the tube length ratio of $\frac{L_t}{L_w}=0.1$. Since the wave lengths are greater than the tube length, the waves complete their cycles more quickly in space, and the phase difference between the rigid and elastic solutions decreases. For the rigid tub case, the waves complete their cycle at an almost infinite speed, and so the phase difference between the rigid and elastic tube solutions becomes less important. It can also be seen that for small Wo the fluid reaches a steady state at all times before the pressure changes, and therefore a quasi-steady state is reached, and, correspondingly, a Poiseuille velocity profile.
\begin{figure}
\centering
\mbox{
\subfigure{\includegraphics[width=\textwidth]{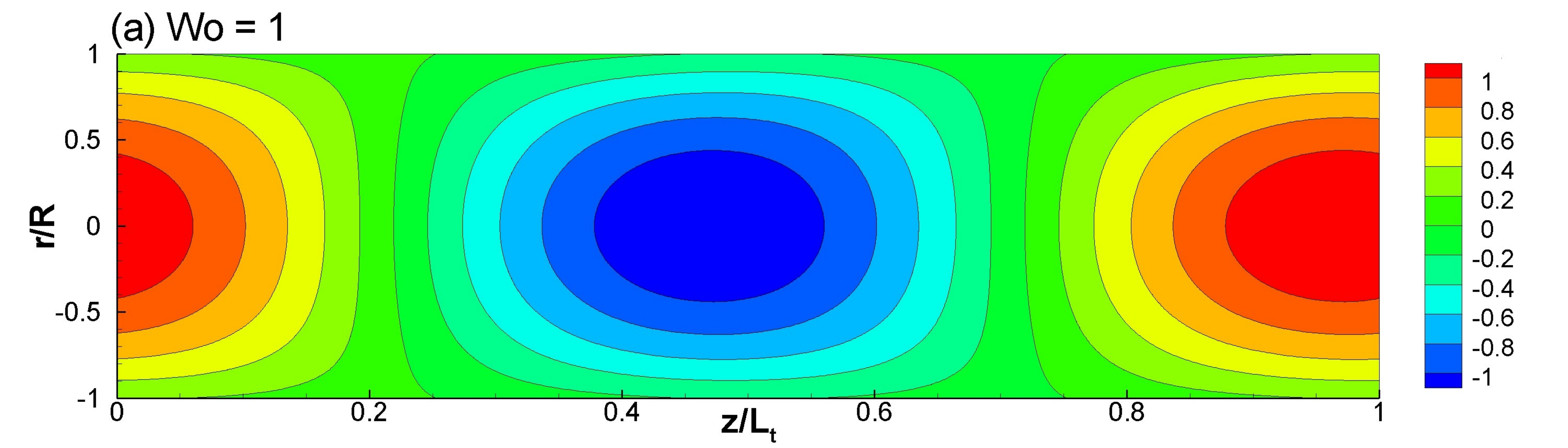}}
}
\mbox{
\subfigure{\includegraphics[width=\textwidth]{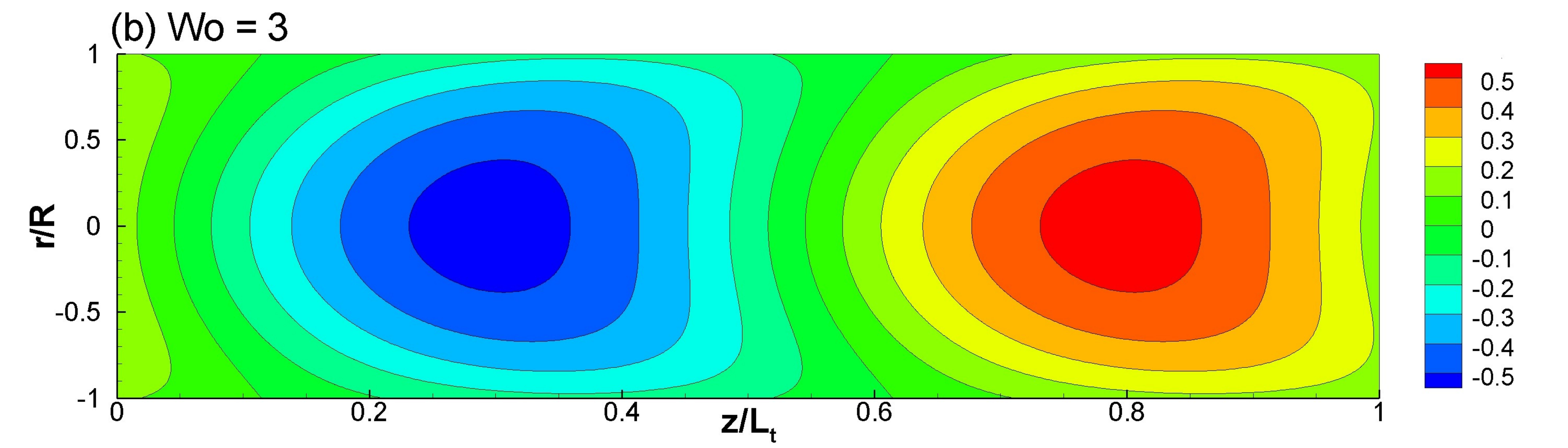}}
}
\caption{Normalized axial velocity contours inside a tube with $L_t/L_w=1$ for (a) $Wo=1$, and (b) $Wo=3$.}
\label{fig:c-a}
\end{figure}
\begin{figure}
\centering
\mbox{
\subfigure{\includegraphics[width=\textwidth]{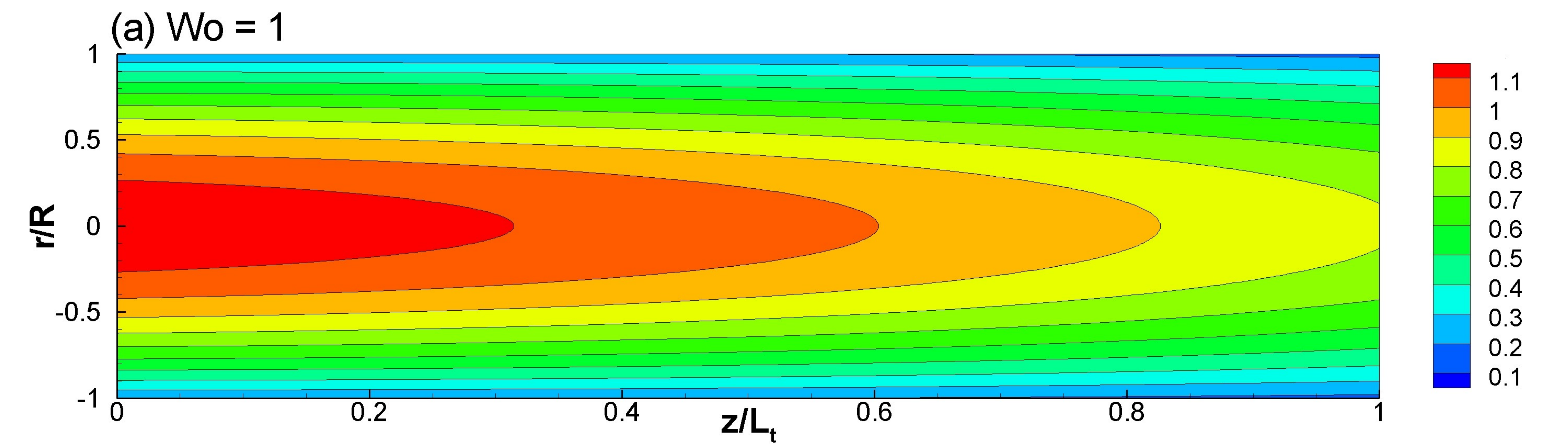}}
}
\mbox{
\subfigure{\includegraphics[width=\textwidth]{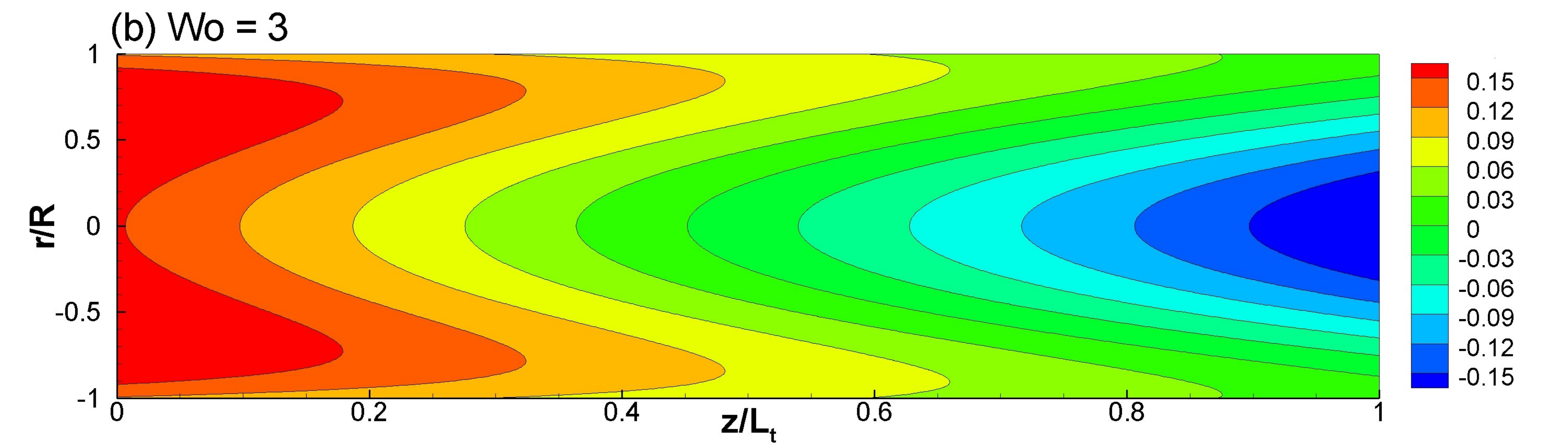}}
}
\caption{Normalized axial velocity contours inside a tube with $L_t/L_w=0.1$ for (a) $Wo=1$, and (b) $Wo=3$.}
\label{fig:c-b}
\end{figure}

Finally, we consider the normalized axial velocity contours for elastic microtubes with two different  tube length to wave length ratios. Fig.~\ref{fig:c-a} and Fig.~\ref{fig:c-b} present the $L_t/L_w=1$ and $L_t/L_w=0.1$ cases, respectively. These figures include axial velocity fields for two different frequency parameters, $\mbox{Wo}=1$, and $\mbox{Wo}=3$, and a slip ratio of $\lambda=0.1$ is used in each of them. Concerning the harmonic oscillations, we plot only one phase state of the axial velocity fields of interest in one time period ($t=0$). As seen in Fig.~\ref{fig:c-a}, the wave completes its cycle in space if the wavelength and tube length are equal. As discussed earlier, the velocity profiles become more parabolic when Wo is small. Increasing Wo affects the flow field by effectively increasing the ratio of oscillatory inertial forces to shear forces. The effect of the slip length as a function of Wo can be seen in the figures, and it is clear that it interacts nonlinearly with the pulsatile flow dynamics.

\section{Conclusions}
\label{sec:con}
The influence of a slip velocity on the flow characteristics in circular elastic microtubes is investigated using a continuum approach with Navier slip boundary conditions. Analytical solutions of the pulsatile flow dynamics in a small tube are found by taking into account the coupling between the fluid motion and the elastic deformation of the tube wall. The formulation derived here can be reduced to the Womersley solution for elastic tubes as a special case by setting the slip length equal to zero. In addition to that, by setting the elasticity factor $G=1$ along with a infinite wave propagation speed, we also recover the classical solution for oscillatory flows through rigid pipes. The influence of slip boundary conditions on the flow through tubes for pulsatile flows is qualitatively different than for steady flow with a constant pressure gradient, wherein slip boundary conditions only result in an added translational effect, without changing the material deformation of the fluid. With an unsteady pressure gradient, however, the influence of slip boundary conditions becomes nonlinear and affects the flow rate, velocity profile, and shear stress, by altering the material deformation in the fluid.

There are large differences between flows in elastic and rigid tubes, especially if the tube length and the wavelength are on the same order. It is already well established that the wall movements in an elastic tube make it easier for the flow to move through the tube than through a comparable rigid tube. Solutions for rigid tubes do not have any wave motion, so velocity profiles at different sections are the same for the same point in the oscillation period (the wave propagation speed $c\rightarrow\infty$). There are wave motions in elastic tubes due to finite wave propagation speeds, however, and this phenomena can be easily seen when the tube length is comparable to the wave length.

In this study, we demonstrated that the presence of slip at the elastic microtube wall enhances flow through the tube. This effect is not just an additional translational velocity, but shows nonlinear behavior and affects the material deformation in the flow field. Additionally, we confirmed that the elasticity of the tube wall also has a positive effect on the flow rate. There is always an increase in flow rate for elastic tubes compared to the inelastic case, and the peak value of the percent change in flow rate increases with increasing slip length. The effective range of the frequency parameter Wo decreases, however, with increasing slip length, and the corresponding value of Wo at this peak value also decreases with increasing slip length. In summary, the presence of velocity slip at the microtube surface has important, nontrivial effects on the velocity profiles. It changes the amount of the phase shift between the fluid inertial motion and the pressure gradient, and alters the material deformation of the media.

\section{References}

\ps Barrat, J.~L. and Bocquet, L.~[1999] ``Large slip effect at a nonwetting fluid-solid interface,"
{\it Physical Review Letters\/} {\bf 82}(23), 4671--4674.

\ps Baudry, J., Charlaix, E., Tonck, A., and Mazuyer, D.~[2001] ``Experimental evidence for a large slip effect at a nonwetting fluid- solid interface,"
{\it Langmuir\/} {\bf 17}, 5232--5236.

\ps Beebe, D.~J., Mensing, G.~A., and Walker, G.~M.~[2002] ``Physics and applications of microfluidics in biology,"
{\it Annual Review of Biomedical Engineering\/} {\bf 4}(1), 261--286.

\ps Cheng, J.~T. and Giordano, N.~[2002] ``Fluid flow through nanometer-scale channels,"
{\it Physical Review E\/} {\bf 65}(3), 312061-312065.

\ps Choi, C.~H., Westin, K.~J.~A., and Breuer, K.~S.~[2003] ``Apparent slip flows in hydrophilic and hydrophobic microchannels,"
{\it Physics of Fluids\/} {\bf 15}(10), 2897-2902.

\ps Chu, A.~K.~H.~[2004] ``Transport control within a microtube,"
{\it Physical Review E\/} {\bf 70}(6), 619021-619025.

\ps Gad-el-Hak, M.~[1999] ``The Fluid Mechanics of Microdevices -- The Freeman Scholar Lecture,"
{\it Journal of Fluids Engineering\/} {\bf 121}, 5--33.

\ps Hansen, J.~S. and Ottesen, J.~T.~[2006] ``Molecular dynamics simulations of oscillatory flows in microfluidic channels,"
{\it Microfluidics and Nanofluidics\/} {\bf 2}(4), 301--307.

\ps Joseph, P. and Tabeling, P.~[2005] ``Direct measurement of the apparent slip length,"
{\it Physical Review E\/} {\bf 71}(3), 353031--353034.

\ps Karniadakis, G., Be{\c{s}}k{\"{o}}k, A., and Aluru, N.~R.~[2005] {\it Microflows and nanoflows: Fundamentals and simulation\/} (Springer Science and Business Media, Inc., New York).

\ps Majumder, M., Chopra, N., Andrews, R., and Hinds, B.~J.~[2005] ``Nanoscale hydrodynamics: Enhanced flow in carbon nanotubes,"
{\it Nature\/} {\bf 438}, 44.

\ps Matthews, M.~T. and Hill, J.~M.~[2007] ``Newtonian flow with nonlinear Navier boundary condition,"
{\it Acta Mechanica\/} {\bf 191}(3), 195--217.

\ps Neto, C., Evans, D.~R., Bonaccurso, E., Butt, H.~J., and Craig, V.~S.~J.~[2005] ``Boundary slip in Newtonian liquids: A review of experimental studies,"
{\it Reports on Progress in Physics\/} {\bf 68}(3), 2859--2897.

\ps Ou, J., Perot, B., and Rothstein, J.~P.~[2004] ``Laminar drag reduction in microchannels using ultrahydrophobic surfaces,"
{\it Physics of Fluids\/} {\bf 16}(12), 4635--4643.

\ps Rothstein, J.~P.~[2010] ``Slip on superhydrophobic surfaces,"
{\it Annual Review of Fluid Mechanics\/} {\bf 42}, 89--109.

\ps Sbragaglia, M. and Prosperetti, A.~[2007] ``Effective velocity boundary condition at a mixed slip surface,"
{\it Journal of Fluid Mechanics\/} {\bf 578}, 435--451.

\ps Squires, T.~M. and Quake, S.~R.~[2005] ``Microfluidics: Fluid physics at the nanoliter scale,"
{\it Reviews of Modern Physics\/} {\bf 77}(3), 977--1026.

\ps Thompson, P.~A. and Robbins, M.~O.~[1990] ``Shear flow near solids: Epitaxial order and flow boundary conditions,"
{\it Physical Review A\/} {\bf 41}(12), 6830--6837.

\ps Thompson, P.~A. and Troian, S.~M.~[1997] ``A general boundary condition for liquid flow at solid surfaces,"
{\it Nature\/} {\bf 389}, 360--362.

\ps Tretheway, D.~C. and Meinhart, C.~D.~[2002] ``Apparent fluid slip at hydrophobic microchannel walls,"
{\it Physics of Fluids\/} {\bf 14}(3), L9-L12.

\ps Vedel, S., Olesen, L.~H., and Bruus, H.~[2010] ``Pulsatile microfluidics as an analytical tool for determining the dynamic characteristics of microfluidic systems,"
{\it Journal of Micromechanics and Microengineering\/} {\bf 20}, 1--11.

\ps Watanabe, K., Udagawa, Y., and Udagawa, H.~[1999] ``Drag reduction of Newtonian fluid in a circular pipe with a highly water-repellent wall,"
{\it Journal of Fluid Mechanics\/} {\bf 381}, 225--238.

\ps Westneat, M.~W., Betz, O., Blob, R.~W., Fezzaa, K., Cooper, W.~J., and Lee, W.~K.~[2003] ``Tracheal respiration in insects visualized with synchrotron X-ray imaging,"
{\it Science\/} {\bf 299}, 558--560.

\ps Whitby, M. and Quirke, N.~[2007] ``Fluid flow in carbon nanotubes and nanopipes,"
{\it Nature Nanotechnology\/} {\bf 2}(2), 87--94.

\ps Womersley, J.~R.~[1955] ``Oscillatory motion of a viscous liquid in a thin-walled elastic tube,"
{\it Philosophical Magazine\/} {\bf 46}, 199--221.

\ps Wu, Y.~H., Wiwatanapataphee, B., and Hu, M.~[2008] ``Pressure-driven transient flows of Newtonian fluids through microtubes with slip boundary,"
{\it Physica A\/} {\bf 387}(24), 5979--5990.

\ps Yousif, H.~A. and Melka, R.~[1997] ``Bessel function of the first kind with complex argument,"
{\it Computer Physics Communications\/} {\bf 106}(3), 199--206.

\ps Zamir, M.~[2000] {\it The physics of pulsatile flow\/} (Springer-Verlag Inc., New York).

\end{document}